\documentclass[final,5p,times,twocolumn,authoryear]{elsarticle}

\usepackage{amssymb}
\usepackage{lipsum}
\usepackage{url}
\usepackage{hyperref}
\usepackage{rotating}
\usepackage{tabularx}
\usepackage{amsmath}

\journal{Astronomy $\&$ Computing}

\begin{document}

\begin{frontmatter}

\title{Supernova scores for active anomaly detection}

\author[sai,phfmsu]{T.~A.~Semenikhin\corref{cor}}
\cortext[cor]{Corresponding author.}
\ead{semenikhintimofey@gmail.com}
\author[sai,hse]{M.~V.~Kornilov}
\author[sai]{M.~V.~Pruzhinskaya}
\author[urfu]{V.~V.~Krushinsky}
% \author[sai]{A.~D.~Lavrukhina}
% \author[clermont]{E.~Gangler}
% \author[clermont]{E.~E.~O.~Ishida}
% \author[ind]{V.~S.~Korolev}
\author[mcwilliams]{K.~L.~Malanchev}
\author[sai]{A.~V.~Dodin}
% \author[iki,sai]{A.~A.~Volnova}
% \author[surrey]{S.~Sreejith}
% \author[]{(The SNAD team)}

\affiliation[sai]{
            organization={Lomonosov Moscow State University, Sternberg astronomical institute},
            addressline={Universitetsky pr.~13}, 
            city={Moscow},
            postcode={119234}, 
            %state={},
            country={Russia}}

\affiliation[phfmsu]{
            organization={Lomonosov Moscow State University, Faculty of Physics},
            addressline={Leninskie Gory~1-2},
            city={Moscow},
            postcode={119991},
            %state={},
            country={Russia}
}

\affiliation[hse]{
            organization={National Research University Higher School of Economics},
            addressline={21/4 Staraya Basmannaya Ulitsa}, 
            city={Moscow},
            postcode={105066}, 
            %state={},
            country={Russia}}

\affiliation[urfu]{organization={Laboratory of Astrochemical Research, Ural Federal University},
addressline = {Ekaterinburg, ul. Mira d. 19},
city = {Yekaterinburg},
postcode={620002},
country={Russia}}

\affiliation[mcwilliams]{
            organization={McWilliams Center for Cosmology and Astrophysics, Department of Physics, Carnegie Mellon University},
            %addressline={}, 
            city={Pittsburgh},
            postcode={PA 15213}, 
            %state={},
            country={USA}}

\begin{abstract}
Large time-domain sky surveys generate extensive multi-year catalogs of light curves in which scientifically valuable transients, such as supernovae (SNe), are vastly outnumbered by artifacts and routine star variability. While supervised machine learning models can efficiently filter known classes, they struggle with extreme class imbalance and may overlook rare or novel events. Conversely, unsupervised anomaly detection provides broad discovery potential but lacks targeted sensitivity. We present a hybrid strategy that integrates a supervised SN probability score (SN-score) into the \texttt{PineForest} active anomaly detection framework to enhance SN discovery rate in the 23rd data release of the Zwicky Transient Facility. We train a binary classifier using light-curve features of spectroscopically confirmed SNe from the ZTF Bright Transient Survey, achieving a $\mathrm{ROC–AUC} \approx 0.98$. Incorporating the SN-score as an additional feature, together with a small set of labeled priors, significantly accelerates the discovery of SN-like transients across ten extragalactic ZTF fields. This method increases discovery efficiency without compromising the ability to identify diverse astrophysical anomalies. Application of the combined methodology resulted in the discovery of seven previously unreported SN candidates, one AGN candidate, one unusual Galactic variable star SNAD283, as well as two host galaxies exhibiting multiple supernova events. These results demonstrate its value for scalable and expert-guided transient search in current and future surveys, including the Vera C. Rubin Observatory Legacy Survey of Space and Time.
\end{abstract}


\end{frontmatter}

\section{Introduction}
\label{introduction}
Time–domain astronomy has entered an era of unprecedented data volume. Modern large-scale sky surveys, such as the Zwicky Transient Facility (ZTF,~\citealt{2019PASP..131a8002B}) and the forthcoming Vera C. Rubin Observatory Legacy Survey of Space and Time (LSST,~\citealt{2009arXiv0912.0201L}), generate up to hundreds of thousands to millions of alerts per night, revealing diverse classes of variable and transient astrophysical phenomena. At these scales, identifying rare and scientifically valuable transients becomes extremely challenging. Many detections correspond to instrumental artifacts, spurious signals induced by atmospheric or observational conditions, or objects of non-astrophysical nature~\citep{2019PASP..131a8003M,Mahabal_2019,2021MNRAS.502.5147M,karpov2023}. Even among real signals, only a small subset may be relevant for a specific scientific objective, while the rest constitute background variability of little interest to a given research program. This imbalance between data volume and scientific relevance motivates the development of efficient automatic methods capable of distinguishing rare, high-value transient events from the overwhelming population of routine detections.

Traditional transient-classification pipelines rely on supervised machine-learning models trained on labelled datasets~\citep{Richards_2011,Bloom_2012,Wright_2015}.
While such models have achieved high performance in filtering astrophysical events from artifacts~\citep{2002SPIE.4836...61A,2007ApJ...665.1246B,Duev_2019} and discriminating between transient types, they inherently struggle with class imbalance and cannot naturally adapt to new or rare phenomena. At the same time, purely unsupervised anomaly detection methods have emerged as powerful tools for discovering unexpected events in large unlabeled datasets~\citep{2019MNRAS.489.3591P,webb2020,2021MNRAS.502.5147M,Villar_2021}. However, these approaches often lack sensitivity to specific object classes of interest unless explicitly guided~\citep{majumder2024,pruzh2023}.

Active anomaly detection frameworks combine the strengths of unsupervised learning with expert feedback, enabling iterative exploration of rare events in vast astronomical datasets~(e.g.~\citealt{ishida2021,KORNILOV2025100960,2025arXiv250503509G}). By integrating domain knowledge in the form of human labels or prior examples, these methods provide a flexible and scalable strategy for targeted discovery while retaining the ability to detect other astrophysical outliers. In practice, a key challenge lies in incorporating available information about known rare classes — such as SNe — without compromising the algorithm's potential to uncover genuinely unexpected objects.

In this work, we introduce the hybrid methodology to perform a supernova (SN) search.
% introduce a  SN probability score  (SN-score) designed to enhance active anomaly detection in ZTF data. 
We develop a supervised binary classifier trained to distinguish SNe from non-SNe using a list of SNe from the Zwicky Transient Facility Bright Transient Survey (ZTF BTS, \citealt{Perley_2020}) cross-matched with ZTF Data Release 23 (DR23). The resulting SN probability score (SN-score), computed for nearly four million objects across ten extragalactic fields in ZTF DR23, is then  employed as an additional informative feature within the \texttt{PineForest} active anomaly detection framework (Ishida et al., in prep.). The purpose of this hybrid strategy is two-fold: (1) to enable the anomaly detection algorithm to autonomously determine a SN detection threshold for particular feature subspace, and (2) to accelerate the identification of transient events with SN-like photometric behavior, without precluding the discovery of other astrophysically interesting anomalies.

% The obtained SN probability score (SN-score) is then employed as an additional informative feature within the \texttt{PineForest} active anomaly detection framework (Ishida et al., in prep.). The purpose of this hybrid strategy is two-fold: (1) to enable the anomaly detection algorithm to autonomously determine a SN detection threshold for particular feature subspace, and (2) to accelerate the identification of transient events with SN-like photometric behavior, without precluding the discovery of other astrophysically interesting anomalies.

% We evaluate the proposed method on almost 4 millions objects across ten extragalactic fields from ZTF DR23. The resulting system demonstrates statistically significant improvements in SN discovery effeciency when guided by both the classifier and a small set of known priors. Importantly, this approach led to the discovery of previously unreported SNe in ZTF DR23, underscoring its practical utility for real survey operations. Our results highlight the potential of integrating supervised classification signals into active anomaly detection pipelines as a scalable strategy for targeted scientific discovery in the era of large synoptic surveys.

The article is organized as follows. In Section~\ref{sec:data}, we describe the datasets used in this work. Section~\ref{sec:preprocessing} outlines the preprocessing pipeline. In Section~\ref{sec:classifier}, we introduce the binary SN classifier. Section~\ref{sec:anomaly_detection} presents the active anomaly detection framework and examines the impact of incorporating the proposed SN-score. The main findings are described in Section~\ref{sec:results}. Conclusions are given in Section~\ref{sec:conclusions}.  Finally, the Appendix~\ref{app} provides the
full list of features and discusses the feature importance. 

\section{Data}\label{sec:data}

\subsection{Zwicky Transient Facility}\label{sec:ztf}
In this work we used data from the Zwicky Transient Facility\footnote{\url{https://www.ztf.caltech.edu/}}  wide-field sky survey. ZTF is conducted at the Palomar Observatory in California, USA, using a 1.26-m Samuel Oschin telescope with a field of view of approximately $47$ square degrees. The survey operates in three photometric passbands ($zg$, $zr$, $zi$,~\citealt{2019PASP..131a8002B}). Data from ZTF are provided in two formats: alerts and data releases. Alerts are issued in real-time, while data releases contain light curves of variable objects over the entire observation period. In this work, we use the $zr$-band light curves from the DR23 for classification and anomaly detection part; light curves in all available passbands for further analysis of SN candidates. ZTF DR23 contains approximately $4.97$ billion light curves. By applying the following selection criteria to all $zr$-band light curves from ZTF DR23: $58178 \le \mathrm{MJD} \le 60125$ and number of good observations per light curve $\geq100$ (this condition leads to the fact that we consider the light curves of only the ZTF primary tiling grid, see Section~2.9 in~\citealt{Bellm_2019}), we obtained a \textit{main} dataset comprising about $720$ million light curves. 

It is important to emphasize that we did not perform any cross-matching between the light curves in a whole main dataset. Consequently, multiple light curves in this dataset may correspond to the same astrophysical source on the sky. We discuss the causes of this in more detail in Section~\ref{sec:cross-match}, but here we note that within the framework of the main dataset this is not critical for us, since we do not train any models sensitive to such duplicates on it entirely.
 
\begin{figure*}[t]
	\centering 
	\includegraphics[width=1\linewidth]{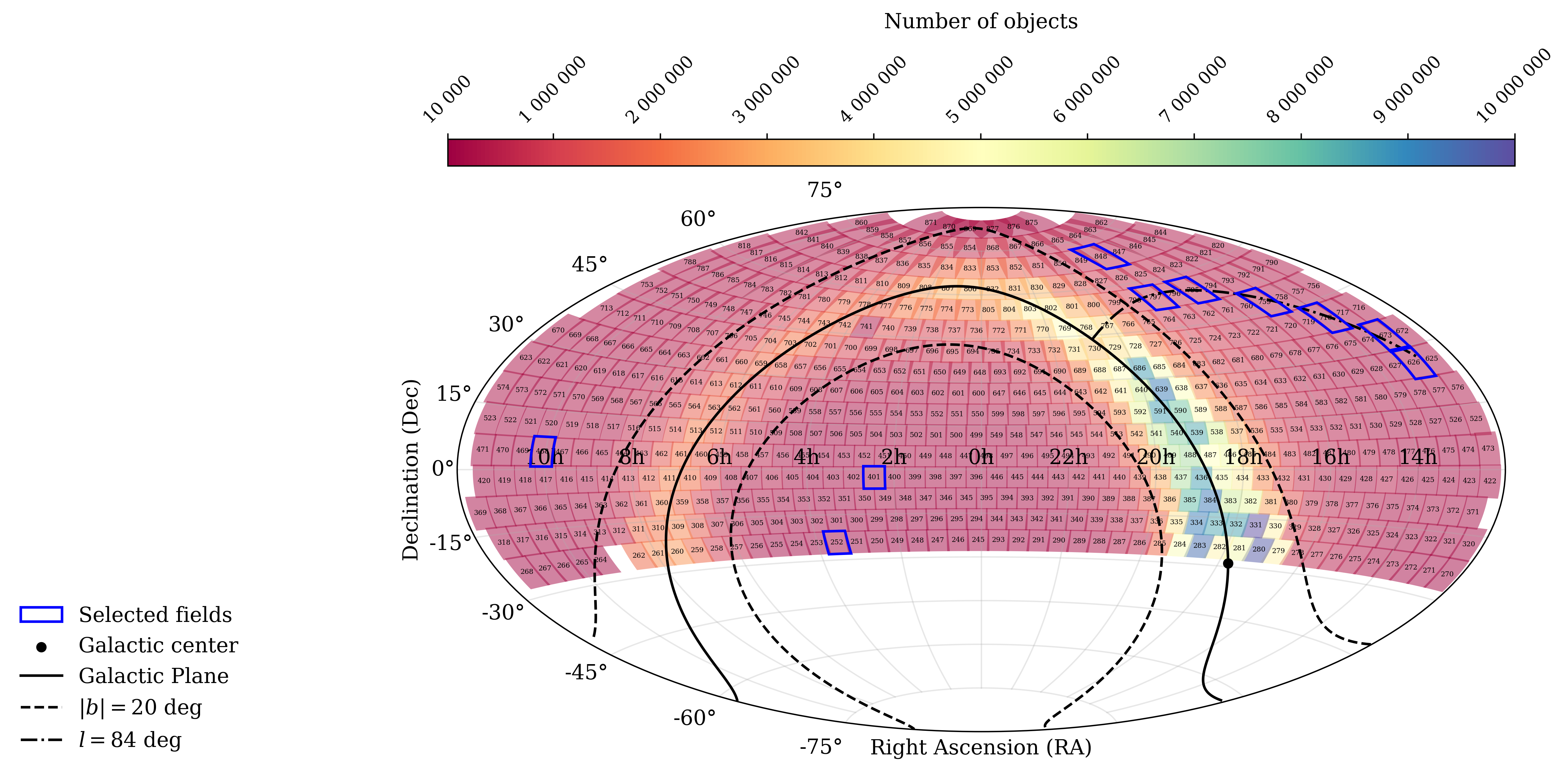}\\[0.5em]
    \includegraphics[width=1\linewidth]{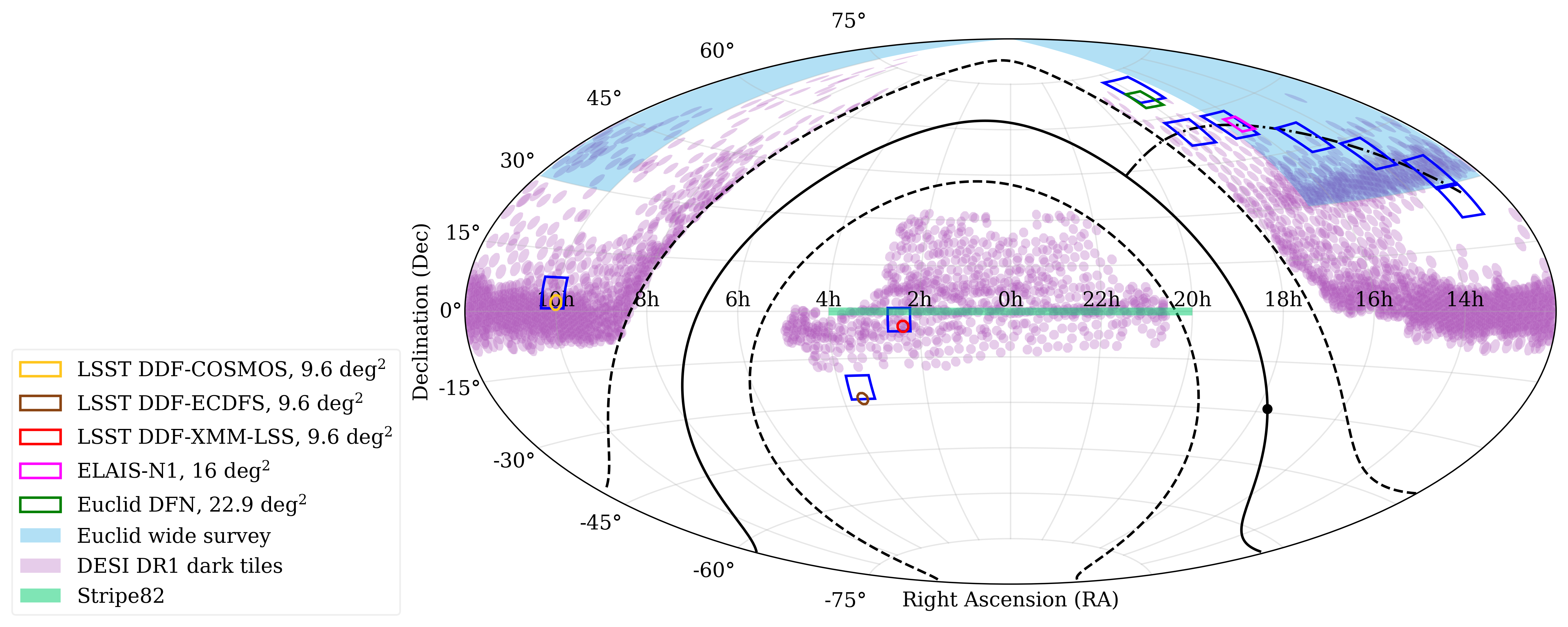}
	\caption{Sky maps illustrating the selection of ZTF fields used in this study. \textbf{Top:} Distribution of object counts from the main dataset across ZTF fields (numbers within rectangles indicate the corresponding {\tt Field ID}); blue rectangles mark the ten selected fields. \textbf{Bottom:} The same ten ZTF fields shown in relation to the sky coverage of other big surveys.}
 
	\label{fig:map}
\end{figure*}

In the ZTF DRs, the entire northern sky is divided into fields identified by specific \texttt{Field ID} numbers. The top part in Fig.~\ref{fig:map} presents a heatmap showing the distribution of object counts across these fields for our main dataset. Since the main dataset contains an extremely large number of objects, it is not feasible to perform active anomaly detection experiments on the full set. Therefore, we limited our analysis to a subset of 10 fields.

The selection of these fields was guided by several criteria: (1) the fields are located outside the Galactic plane ($|b| > 20$ deg); (2) the fields overlap with other survey projects (see the bottom part in Fig.~\ref{fig:map}); and (3) 
several fields are elongated along the same Galactic longitude in order to account for population changes of Galactic anomalies with Galactic latitude. Considering these factors, we selected the fields indicated by blue rectangles in Fig.~\ref{fig:map} for the active anomaly detection experiments.

\subsection{Bright Transient Survey}\label{BTS}
To construct a classifier distinguishing SNe from non-SNe, we require a sample of confirmed SNe with corresponding observations available in ZTF DR23. For this purpose, we used data from the ZTF BTS.
 
The ZTF BTS is a large spectroscopic SN survey that has operated since 2018, targeting nearly all extragalactic, non-AGN transients brighter than 18.5--19 mag across the northern sky. As of 2025, it includes over 11,000 spectroscopically confirmed transients, the vast majority of which are SNe. BTS achieves a spectroscopic completeness of approximately 95\% for well-sampled, SN-like light curves, with most of the remaining incompleteness due to weather or operational gaps. The survey applies well-defined selection criteria to construct a high-quality statistical sample and provides subtype information when available. Although the main survey phase has ended, BTS continues to classify all bright transients and those in nearby galaxies down to fainter limits, ensuring a nearly complete and uniform sample of bright SNe from ZTF.

We selected all classified SNe from the BTS, then retained only those that satisfy the BTS quality criteria, and further restricted the sample to objects whose peak end date fall within $58178 \le \mathrm{MJD} \le 60125$. As a result, out of approximately 11,000 SNe in BTS, the sample of 4,832 objects remained.

\section{Preprocessing}\label{sec:preprocessing}

\subsection{Feature extraction}
In machine learning, before applying a specific algorithm to the data, it is often necessary to perform an intermediate step -- feature extraction. In our case, this step is essential, since the light curves provided in the ZTF DRs are irregularly sampled in time and vary in length. To convert the data into a tabular format suitable for analysis, we used the Python library \texttt{light-curve}\footnote{\url{https://github.com/light-curve/light-curve-python}}~\citep{2021MNRAS.502.5147M}, which enables the extraction of informative astronomical features from light curves.

The set of 47 extracted features is hereafter referred to as the \textit{initial} feature set. The complete list of these features is provided in~\ref{app:features}. We extracted features for all light curves in the main dataset using 14 cores of a dual Intel Xeon Gold 5118 system, which took approximately 235 hours. The resulting file contains about 720 million rows and has a total size of 127 GB.

\subsection{Cross-matching}\label{sec:cross-match}
The BTS sample provides sky coordinates for each object, which allowed us to cross-match these SNe with objects from our main dataset. This was done using an application programming interface\footnote{\url{https://db.ztf.snad.space/api/v3/help}} that, given celestial coordinates and a search radius, returns all object IDs (OIDs) from ZTF DR23 located within the specified area. Using this tool, we retrieved all objects from the main sample lying within a 2-arcsecond radius centered on each BTS supenova. In cases where multiple matches were found within this radius, we selected the object with the largest number of observations in the $zr$ filter. This choice is motivated by the fact that, in ZTF, a single astrophysical source may be assigned different OIDs when it falls near the boundary between two fields, resulting in its light curve being split across multiple OIDs. Since these OIDs correspond to the same physical source, we adopt the OID with the greatest number of
$zr$-band observations, assuming it provides the most complete and temporally well-sampled representation of the underlying light curve. As a result of this cross-matching procedure, we obtained a list of 2,066 objects from our main sample those overlap with BTS sample. After visual inspection of all candidates, the final sample contained 674 SNe.

\section{Binary SN classifier}\label{sec:classifier}
We employed a standard supervised classification approach, in which each object is represented by a set of initial features. During training, these features are provided to the model along with class labels, allowing it to learn a mapping between feature space and object type. During inference, the model receives only the feature vectors of unlabeled objects and outputs a continuous SN-score that can be interpreted as the probability of the object being a SN according to the classifier.

\subsection{Train dataset}\label{sec:clf_train}
Since the fraction of SNe in the main dataset is extremely small, it was not feasible to develop the classifier using a balanced training sample (see~\ref{app:selection}). For training, we assigned positive class to all 674 confirmed SNe (see Section~\ref{BTS}), while the negative class included $6300$ non-SN objects from the Anomaly Knowledge Base (AKB; see Section 4 in \citealt{2023PASP..135b4503M}) together with 10,000 randomly selected objects from the main dataset. This approach makes the classifier more conservative, reducing the frequency of high SN-scores and thereby improving its overall precision.

\subsection{Model}
For the binary classifier, we chose a Random Forest~\citep{rand} model, as previous studies~\citep{SEMENIKHIN2025100919} have shown that popular machine learning methods achieve comparable performance on similar tasks. As described in Section~\ref{sec:clf_train}, we used an unbalanced SN sample and additionally assigned class weights of 1:1000 in the Random Forest to prevent the model from ignoring the rare class. Otherwise, we used the default parameters from the \texttt{scikit-learn}\footnote{\url{https://scikit-learn.org/stable/}} implementation (number of trees -- 100, the minimum number of samples required to split an internal node -- 2, criterion -- Gini).

\begin{figure}[t]
	\centering
    \includegraphics[width=1\linewidth]{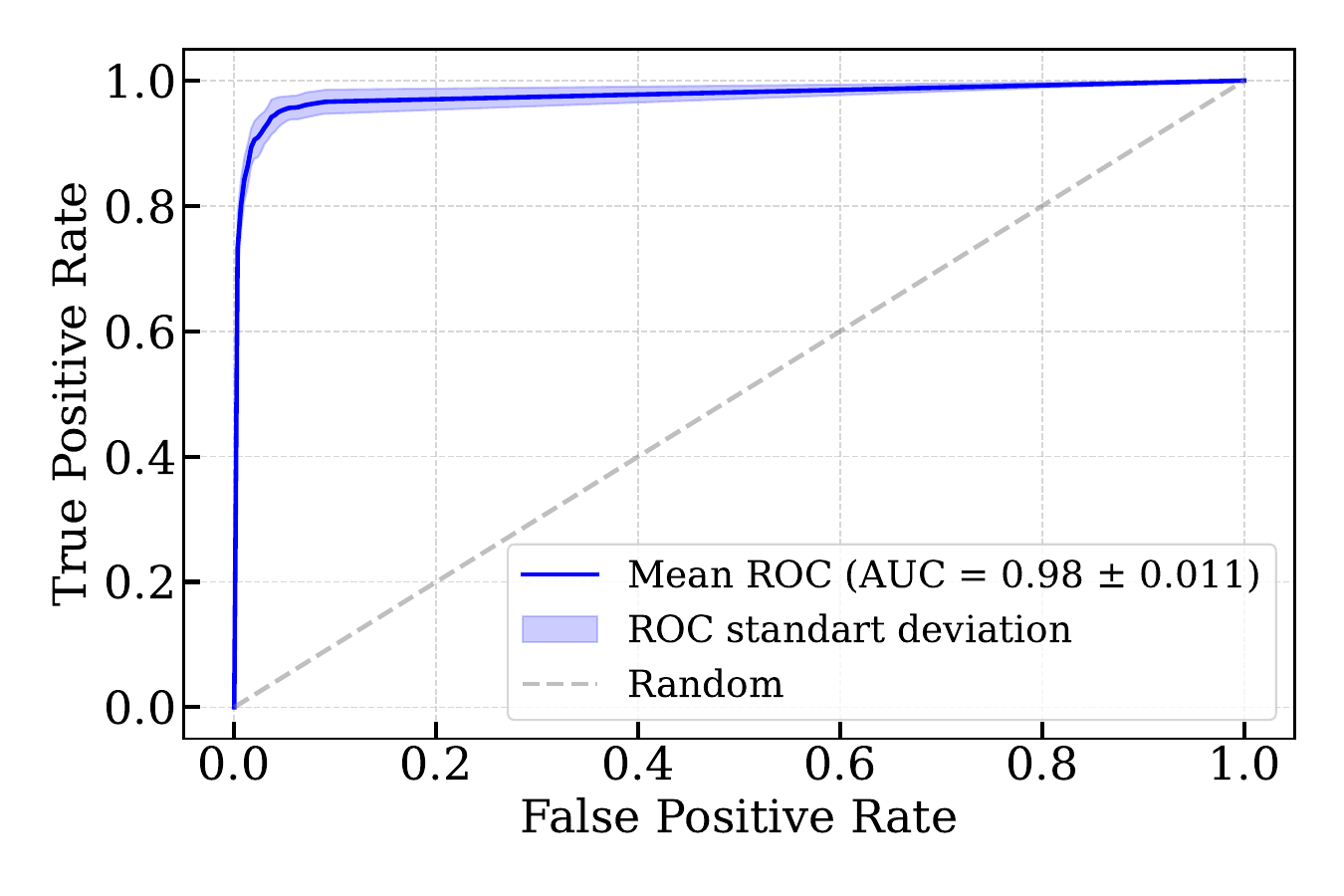}
	\caption{Mean ROC curve across stratified k-fold validation folds, with the shaded band indicating $\pm \sigma$ variability.}
	\label{fig:clf_roc_curves}
\end{figure}

To evaluate the performance of the final binary classifier, we used stratified k-fold ($k = 5$) cross-validation, which accounts for class imbalance when splitting the data into folds. The primary performance metric was the ROC-AUC, which is threshold-independent; the final classifier achieved a $\mathrm{ROC-AUC} = 0.98 \pm 0.011$ (see Fig.~\ref{fig:clf_roc_curves}).

\section{Anomaly detection}\label{sec:anomaly_detection}
The concept of anomaly detection is to identify unusual astrophysical objects within an unlabeled dataset. In our case, we search for such objects within a small subset of the main dataset -- specifically, in the 10 selected fields (see Section~\ref{sec:ztf}). In general, anomaly detection operates by taking objects represented in some feature space (in our case, light-curve features) and feeding them into a model that estimates the density distribution of objects within this space, ranking them from the most to the least common. The main innovation proposed in this work is to incorporate the predictions of a binary SN classifier as an additional feature in the initial feature set, resulting in an \textit{augmented} feature set. This approach allows the anomaly detection method to determine the optimal decision threshold of the classifier autonomously.

Traditional static anomaly detection methods, such as Isolation Forest~\citep{isolationforest}, do not account for the specific preferences of an expert and do not incorporate user feedback. In contrast, active anomaly detection methods allow the model to adapt dynamically to expert input during the exploration process. In this work, we tested the performance of \texttt{PineForest}, an implementation available in the {\tt Python} library \texttt{Coniferest}~\citep{KORNILOV2025100960}.

\subsection{\texttt{PineForest}}
This is an iterative algorithm whose first step is the initialization of an Isolation Forest. The expert is then presented with the most anomalous object in the dataset, which is carefully examined using all available information sources. The expert labels the object as \textbf{A} (anomalous, i.e., scientifically interesting), \textbf{R} (regular, a typical object of no particular interest), or \textbf{U} (unknown, when a confident judgment cannot be made; such objects are excluded from the dataset and do not affect model training). Next, \texttt{PineForest} randomly constructs additional trees and evaluates their performance together with the existing ones in the model, assessing which trees best align with the expert’s feedback and discarding a fraction of those that perform poorly. The remaining trees are retained in the model, and the process repeats iteratively until a predefined review budget is reached.

Despite the adaptive nature of this algorithm, achieving high-quality model performance may sometimes require a large review budget. However, if the expert has prior knowledge of the object type to be discovered and possesses examples of confirmed, classified instances within the dataset, \texttt{PineForest} allows these to be provided as priors. In this case, immediately after the initialization step of the Isolation Forest, \texttt{PineForest} prunes trees that are inconsistent with the supplied priors. The algorithm then proceeds following the standard iterative procedure.
Initial experiments have demonstrated that providing even a small number of priors (as few as ten) can significantly accelerate the anomaly detection process (see Table~\ref{tab:result}).

\subsection{Evaluation}
To evaluate the performance of the proposed approach, we measured how many SNe were recovered by \texttt{PineForest} among the 30 visually inspected candidates in different experiments. We conducted experiments both without and with priors, using the initial and the augmented feature sets. Table~\ref{tab:result} summarizes the results for each field, showing the number of SNe identified among the 30 examined candidates. The last column of the table reports the number of SNe found within the top 30 objects ranked by the binary classifier’s SN-score.

To assess the statistical significance of the observed differences between the experimental setups, we complemented the main results (Table~\ref{tab:result}) with a comparative analysis using Fisher’s exact test (Table~\ref{tab:fisher_results}). Fisher’s exact test evaluates whether two categorical variables are independent by calculating the probability of obtaining the observed contingency table under the null hypothesis. In our case, the null hypothesis ($H_0$) states that both methods under comparison recover SNe with equal efficiency, while the alternative hypothesis ($H_1$) assumes that the second method yields a higher detection rate. The resulting p-value quantifies the probability of obtaining the observed (or more extreme) results under $H_0$; values below 0.05 are typically considered statistically significant.  

For each pair of methods, the contingency tables summarize the number of detected and non-detected SNe across the ten ZTF fields. The Odds Ratio indicates the relative likelihood of successful detection by the first method compared to the second: values below 1 imply that the second method performs better, while values above 1 favor the first method.

The results (Table~\ref{tab:fisher_results}) show that introducing priors in combination with the augmented feature set leads to a highly significant improvement ($p \ll 0.05$), confirming the benefit of guiding the anomaly detection algorithm with even a small number of known examples. In contrast, adding priors to the initial feature set or using the augmented features alone without priors did not produce statistically significant gains. Interestingly, the comparison between the augmented feature set with priors and the simple ranking by the binary classifier’s SN-score reveals no significant difference ($p = 0.50$), indicating that both approaches yield comparable SN recovery rates across the tested fields. This is further supported by a strong Spearman rank correlation ($\rho = 0.82$) between their field-wise detection counts, indicating consistent detection behavior. 
Although these results may appear to lessen the role of active anomaly detection in the specific task of SN recovery, it is important to stress that our broader objective is not to perform a binary SN/non-SN classification (such events are already efficiently identified within the alert stream). Instead, these experiments serve to demonstrate that our approach can effectively highlight the regions of feature space most relevant to the phenomena of interest, while simultaneously making this space more structured and discriminative through the inclusion of an additional informative feature. In future applications, our goal is to target much rarer transients, such as gravitational lenses, and we expect that the proposed approach will help guide the search toward these scientifically valuable outliers.

\begin{table*}
\begin{center}
\begin{tabular}{|c | c c | c c | c |} 
    \hline
     & \multicolumn{2}{ c |}{initial} & \multicolumn{2}{ c |}{augmented} & \\
    \cline{2-5}
    Field ID & without priors & with priors & without priors  & with priors & top30 SN-score \\
    \hline
     252 & \bf{1} &    0   & \bf{1} & \bf{1} & {0} \\
     401 &   1    &    2   &   1 & \bf{4} & {3} \\
     468 &   0    &    0   &   0 & {0}    & \bf{1} \\
     795 &   0    &    2   &   0 & \bf{8} & \bf{8} \\
     848 &   0    &    4   &   0 & \bf{9} & {6} \\
     797 &   0    &    0   &  {0} & \bf{8} &  \bf{8} \\
     759 &   0    &    0   &   0   & 1      & \bf{3} \\
     718 &   2    &    5   &   2   & \bf{8} & \bf{8} \\
     673 &   1    &    0   &   0   & 4      & \bf{6} \\
     626 & \bf{2} &    1   &   0   & \bf{2} & 1 \\
    \hline
\end{tabular}
\caption{Results of the \texttt{PineForest} runs on the selected fields (see Section~\ref{sec:ztf}). The table lists the numbers of SNe identified among top 30 visually inspected candidates. Before performing the anomaly search, all SNe used in the classifier training were removed from the fields. The best result for each field is highlighted in bold.}
\label{tab:result}
\end{center}
\end{table*}

{
\renewcommand{\arraystretch}{1.3}
\begin{table*}[h]
\centering
\begin{tabularx}{\textwidth}{X c c c X}
\hline
Comparison & Contingency table (SNe / non-SNe) & Odds Ratio & p-value & Interpretation \\
\hline
initial without priors  vs. initial with priors  &
\raisebox{-.5\height}{$\begin{bmatrix} 7 & 293 \\ 14 & 286 \end{bmatrix}$} &
0.49 & 0.09 &
Moderate difference, not statistically significant (p $>$ 0.05). Using priors without SN-scores likely improves the result, but evidence is insufficient. \\

augmented without priors  vs. augmented with priors &
\raisebox{-.5\height}{$\begin{bmatrix} 4 & 296 \\ 45 & 255 \end{bmatrix}$} &
0.08 & $10^{-11}$ &
Highly significant difference — adding priors to augmented feature set substantially improves performance (p $ \ll $ 0.05). \\

initial without priors  vs. augmented without priors  &
\raisebox{-.5\height}{$\begin{bmatrix} 7 & 293 \\ 4 & 296 \end{bmatrix}$} &
1.77 & 0.90 &
No significant difference between the initial and augmented feature set when priors are not used. \\

initial with priors  vs. augmented with priors  &
\raisebox{-.5\height}{$\begin{bmatrix} 14 & 286 \\ 45 & 255 \end{bmatrix}$} &
0.28 & $10^{-5}$ &
Adding SN-scores to features significantly improves the results, when we use priors. \\

top30 SN-score vs. augmented with priors  &
\raisebox{-.5\height}{$\begin{bmatrix} 44 & 256 \\ 45 & 255 \end{bmatrix}$} &
0.97 & 0.50 &
No significant difference; both methods show very similar behavior across fields. \\
\hline
\end{tabularx}
\caption{Results of pairwise Fisher’s exact tests between different detection methods. 
The contingency tables show the number of detected versus non-detected SNe for each pair of methods. 
The reported one-sided p-values correspond to Fisher’s exact tests under the hypothesis that the second method performs better than the first.}
\label{tab:fisher_results}
\end{table*}
}

% \begin{figure}[t]
% 	\centering
%     \includegraphics[width=1\linewidth]{figs/correlation.pdf}
% 	\caption{Spearman correlation matrix between the tested methods.
% The plot shows pairwise Spearman correlation coefficients computed over the number of supernovae (SNe) identified in each field for the different search configurations.
% High positive correlations (red) indicate that the corresponding methods tend to detect SNe in the same fields, suggesting similar selection behavior.
% Low or negative correlations (blue) imply that the methods are sensitive to different types of fields or SNe candidates.}
% 	\label{fig:correlation}
% \end{figure}

\section{Discovered transients and special cases in ZTF DR23}
\label{sec:results}

During the development of the binary classifier and the subsequent \texttt{PineForest} experiments, seven previously unreported SN candidates, one AGN candidate and one unusual variable star were identified in ZTF DR23. Although all of these events appeared in the ZTF alert stream (as indicated by their assigned ZTF IDs), but they were not flagged as noteworthy by broker pipelines and therefore escaped earlier attention. All discovered objects have been submitted to the Transient Name Server\footnote{\url{[https://www.wis-tns.org/}} (TNS), and they are listed in Table~\ref{tab:sn_results}. The first column contains the internal SNAD object name; the second -- the corresponding ZTF identifier; the third -- unique ZTF DR OID; the equatorial coordinates in degrees are given in column~4; the peak absolute magnitude in the $zr$ band, $M_{r,\mathrm{max}}$, derived from the best-fitting light-curve model is shown in column~5; and the preferred classification in given in column~6.

{
\renewcommand{\arraystretch}{1.3}
\begin{table*}[h]
\begin{center}
\begin{tabularx}{\textwidth}{
    l
    c
    c
    c
    c
    >{\raggedleft\arraybackslash}X
}
\hline
Name & ZTF ID & OIDs & $\alpha, \delta$ (deg) & $M_{r, max}$ & Classification \\
\hline
SNAD280 & ZTF20aanskdl & 790209200005857 & 195.95520 55.91314 & $-18.62^m$ & Ia \\
SNAD281 & ZTF19aarisyl & 676207400000993 & 219.21985 32.27425 & $-19.95^m$ & IIn \\
SNAD282 & ZTF18aarqjgu & 676212300024137 & 215.84439 33.72295 & $-16.36^m$& Ia \\
SNAD283 & ZTF23aaaptpf & 787210100006486 & 161.37753 56.05923 & -- & VS of MW galaxy \\
SNAD284 & ZTF19aawvklc & 759201400013865 & 229.08777 44.56771 & $-18.87^m$& Ia \\
SNAD285 & ZTF18abmkbqk & 759209300030421 & 227.80917 48.55150 & $-15.67^m$ & Ia\\
SNAD286 & ZTF21acmivtg & 252216200002097 & 49.73880 -20.91431 & $-19.50^m$ & IIP\\
SNAD287 & ZTF19aaltuxk & 533202300011210 & 243.77243 8.83445 & $-19.11^m$ & Ia \\
SNAD288 & ZTF22abeuuvu & 650202200001826 & 17.11930 31.16565 & -- & AGN \\
\hline
\end{tabularx}
\caption{Transients identified during the anomaly detection experiments. The values of $M_{r, max}$ and classification of SN candidates inferred from the \texttt{SNCOSMO} light-curve fits.}
\label{tab:sn_results}
\end{center}
\end{table*}
}

\subsection{Supernova candidates}
\label{sec:new_sne}

We performed an initial light curve analysis of seven SN candidates using the Python package \texttt{SNCOSMO}\footnote{\url{https://sncosmo.readthedocs.io/en/stable/}}. For each candidate, the light curves in all available passbands ($zg$, $zr$, $zi$) were fitted with the Nugent's SN models\footnote{\url{https://c3.lbl.gov/nugent/nugent_templates.html}} and the \texttt{SALT2} model~\citep{2014A&A...568A..22B}. The Nugent's templates are simple spectral time-series parameterized by redshift ($z$), the observer-frame time corresponding to the explosion moment ($t_0$) and the amplitude.

The \texttt{SALT2} model, developed for Type~Ia SNe, describes the spectral energy distribution as a combination of a mean template, a variability component governing light curve shape, and a color law. Its free parameters are stretch parameter ($x_1$), color parameter ($c$), redshift, amplitude and time of maximum light.

\begin{figure*}[t]
\centering

\includegraphics[width=0.48\linewidth]{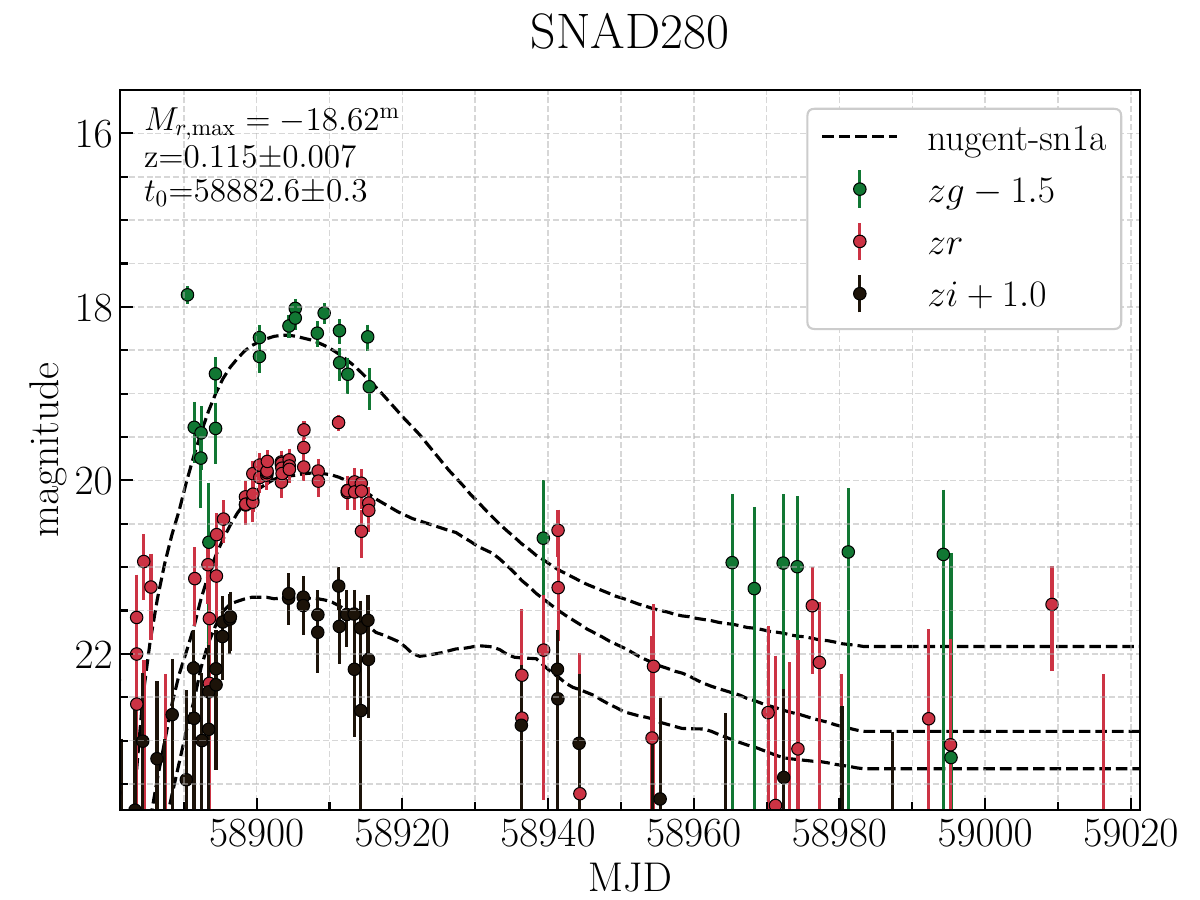}\hfill
\includegraphics[width=0.48\linewidth]{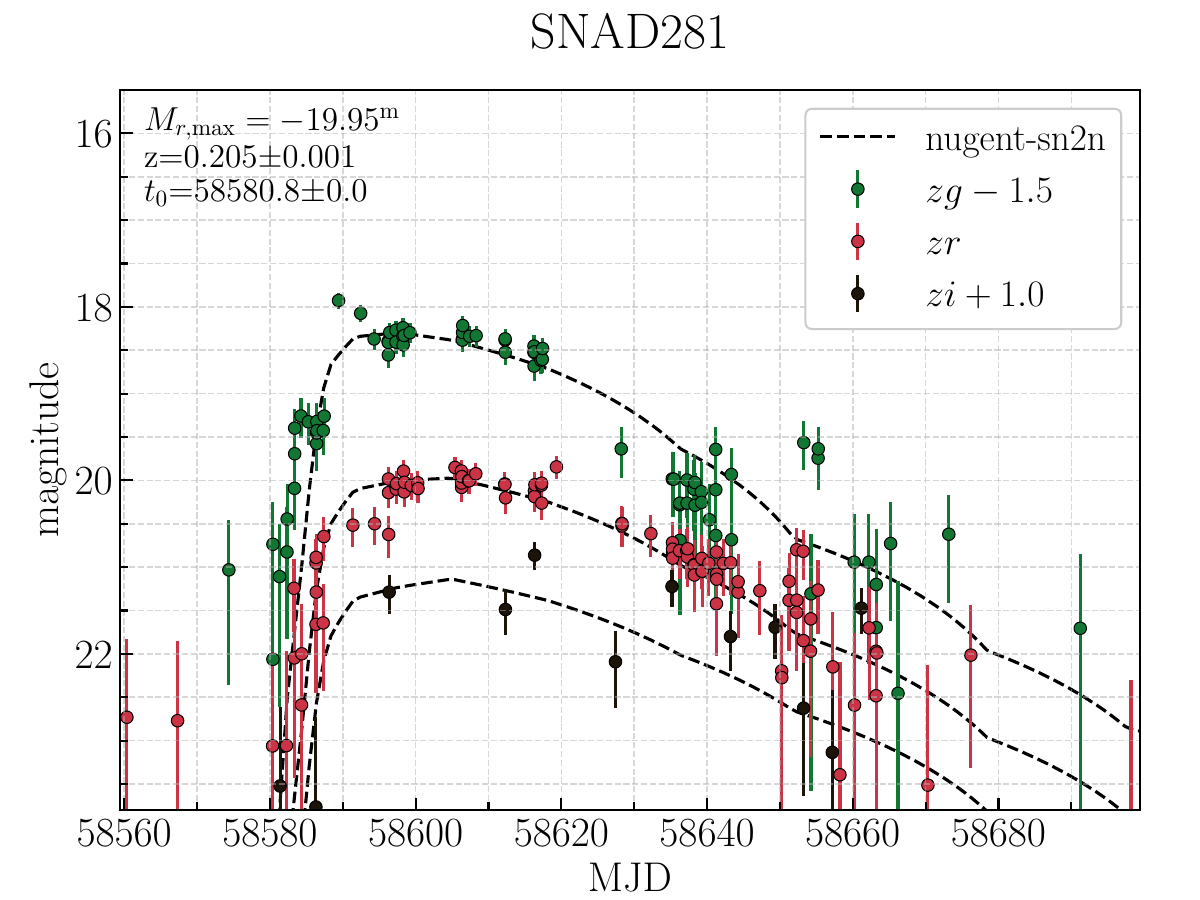}

\includegraphics[width=0.48\linewidth]{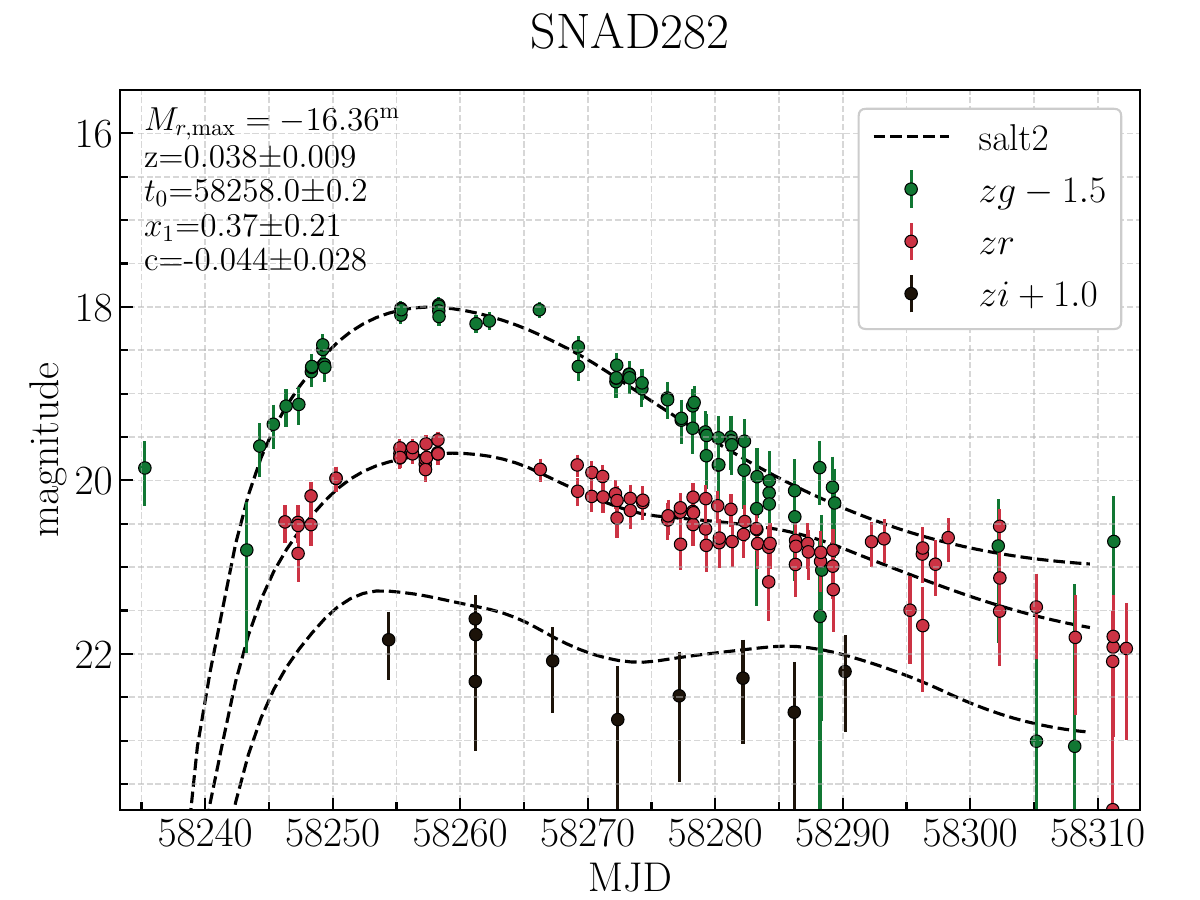}\hfill
\includegraphics[width=0.48\linewidth]{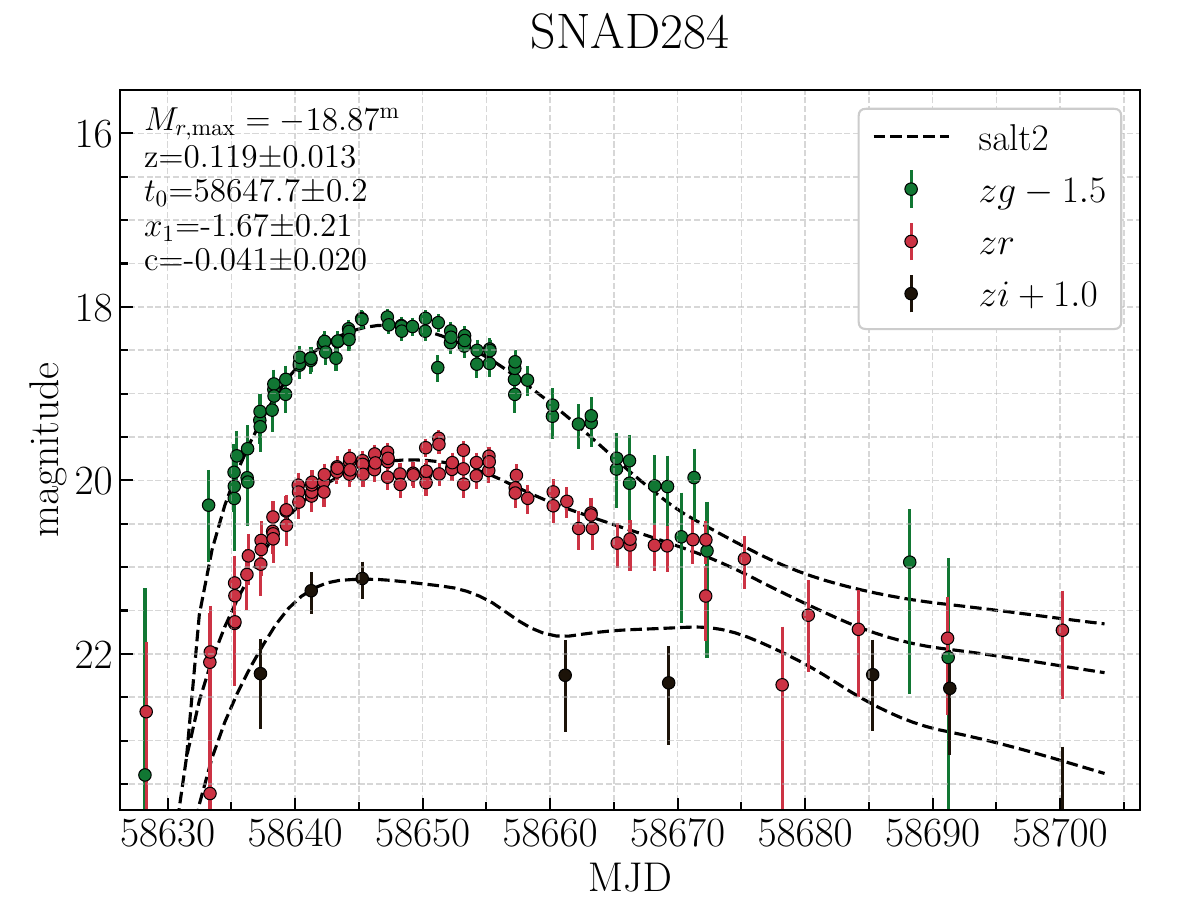}

\includegraphics[width=0.48\linewidth]{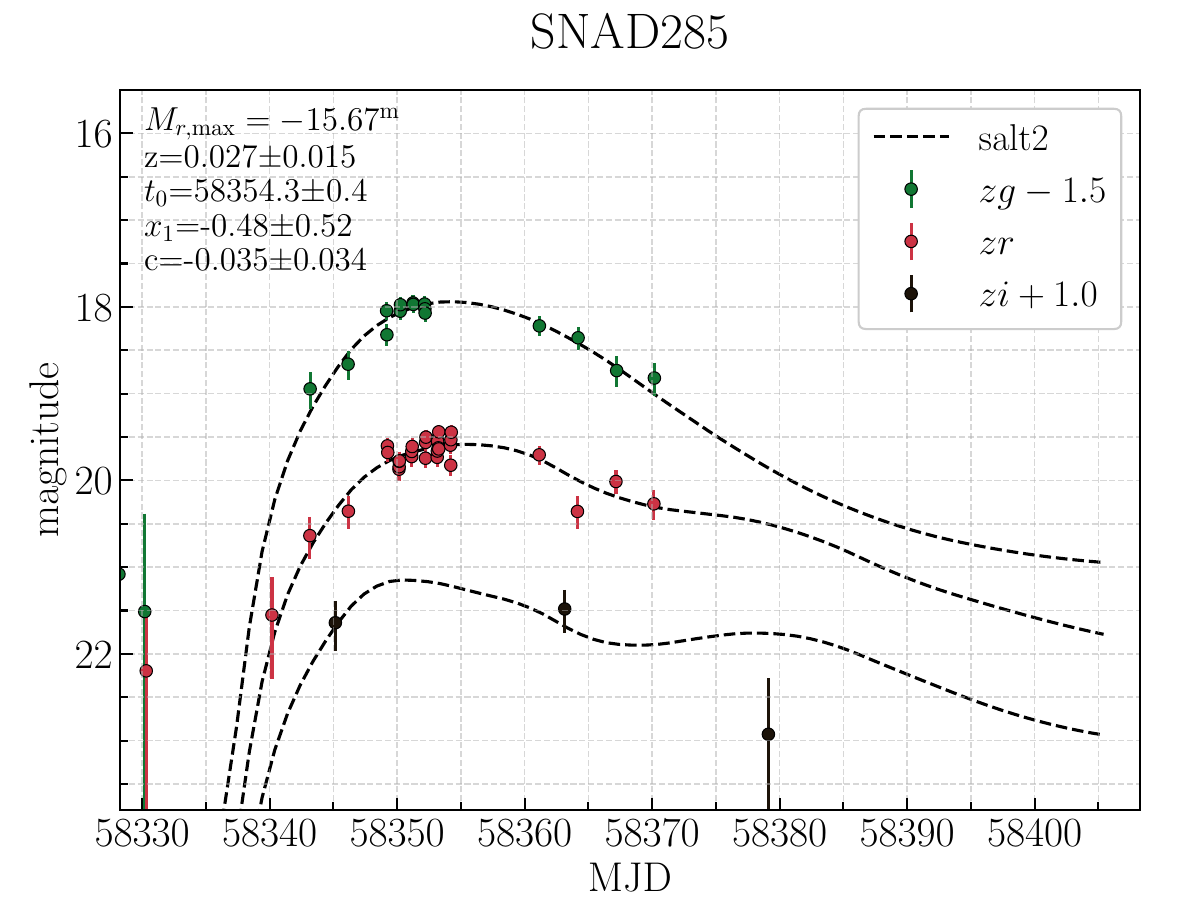}\hfill
\includegraphics[width=0.48\linewidth]{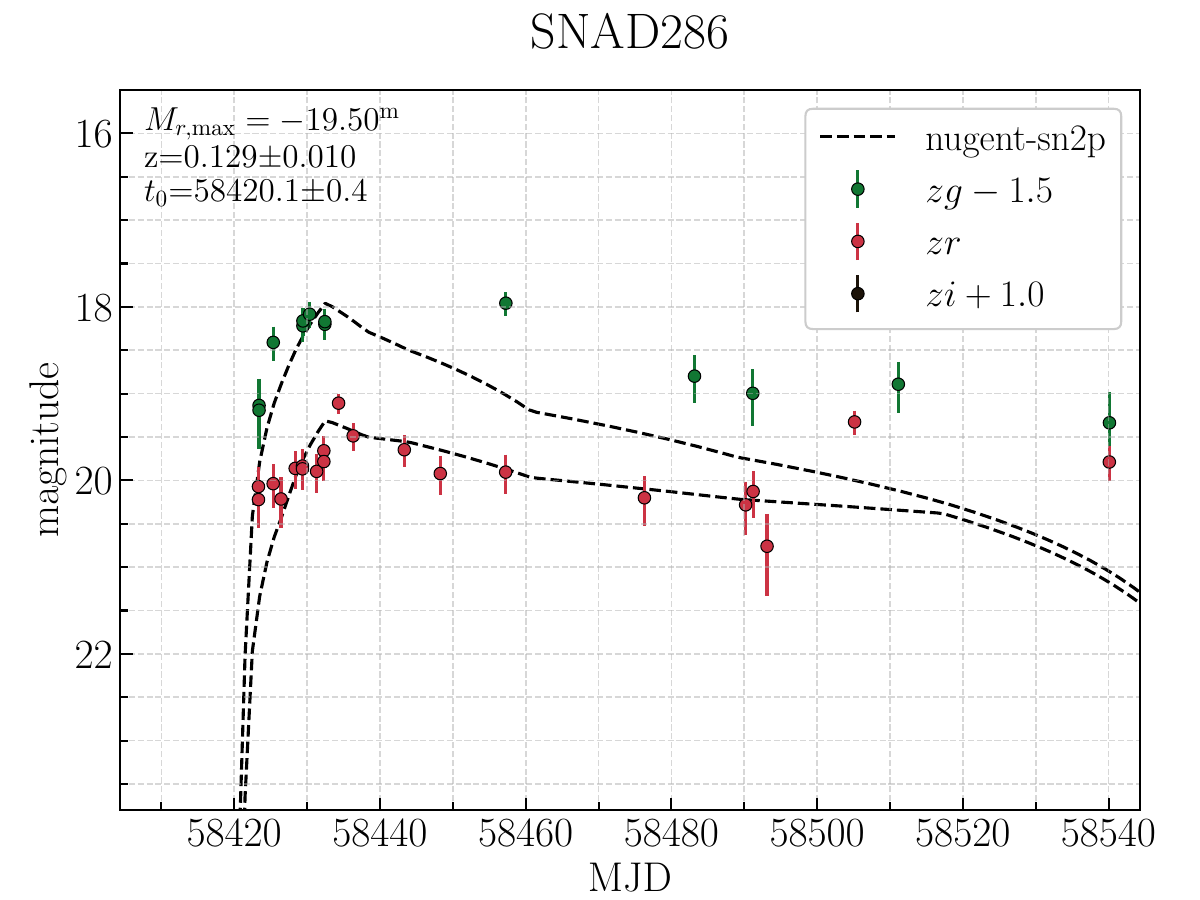}
\caption{ZTF DR23 multicolor light curves of new supernova candidates with \texttt{SNCOSMO} best-model fits and fitted model parameters.}
\label{snad_fits}
\end{figure*}

To mitigate host–galaxy contamination in the ZTF photometry, we subtracted the reference magnitudes provided in the SNAD catalogue\footnote{\url{https://snad.space/catalog/}}. The light curves were further corrected for Galactic extinction using the line-of-sight reddening estimates of~\cite{2011ApJ...737..103S}. For each SNAD SN candidate, the preferred model was selected by minimizing the $\chi^2$ statistic. The light curve fitting results for the discovered SNe are shown in Fig.~\ref{snad_fits} and ~\ref{fig:snad287}, and the optimized model parameters are listed in the corresponding panels. Based on the best-fitting photometric models, our sample consists of five SNe~Ia, one SN~IIn, and one SN~IIP.

\subsection{A Galactic helium-rich transient: SNAD283}

\begin{figure*}
	\centering
    \includegraphics[width=1\linewidth]{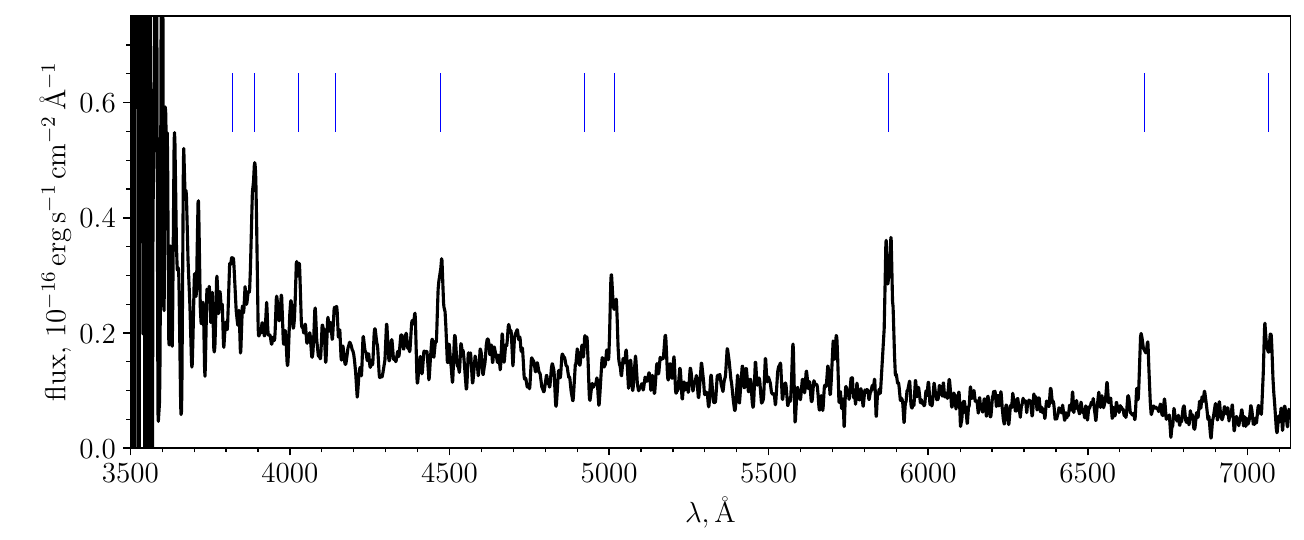}
	\caption{The spectrum of SNAD283 obtained with the 2.5-meter telescope of the Caucasus Mountain Observatory. Blue markers indicate the positions of HeI double-peaked emission lines.}
	\label{fig:snad283_spectrum}
\end{figure*}

SNAD283 was initially identified as a potential superluminous supernova candidate, based on a single SN-like outburst with a broad light curve lasting for more than one year. To determine its redshift and nature, we obtained a spectrum (Fig.~\ref{fig:snad283_spectrum}) with the Transient Double-beam Spectrograph (TDS) mounted on the 2.5-m telescope of the Caucasus Mountain Observatory (CMO; \citealt{2020AstL...46..836P,2020gbar.conf..127S}).

\begin{figure}
	\centering
    \includegraphics[width=1\linewidth]{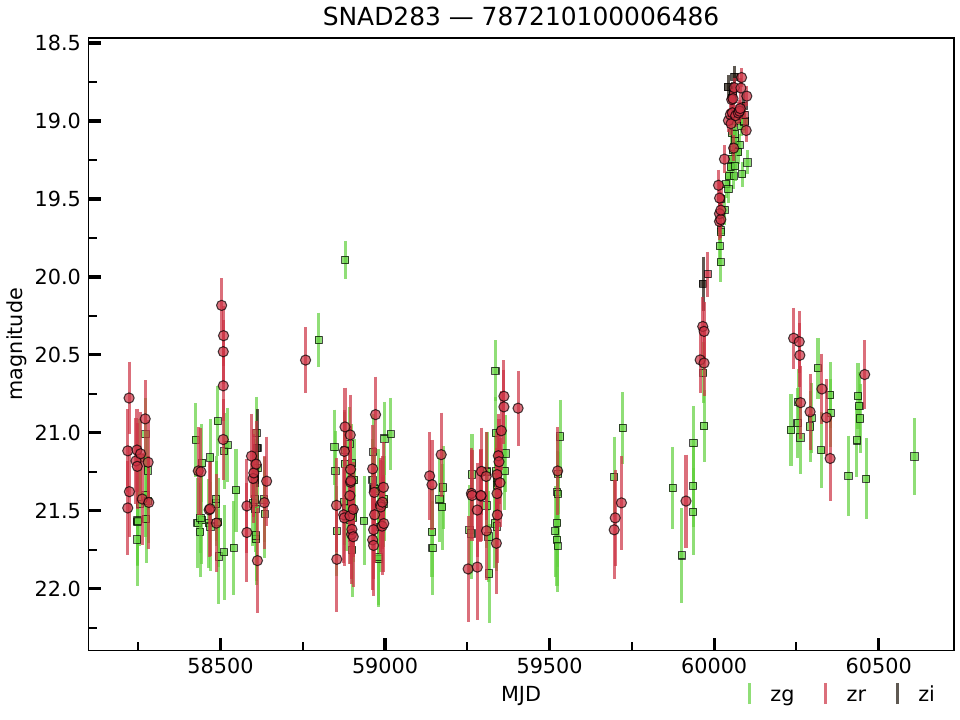}
	\caption{ZTF DR23 multicolor light curve of SNAD283 generated with the SNAD ZTF Viewer \citep{2023PASP..135b4503M}.}
	\label{fig:snad283_lc}
\end{figure}

The spectrum  shows that the source is a Galactic object. It is dominated by strong helium double-peaked emission lines, indicating a helium-rich accretion disc. While the event duration is compatible with that of some classical novae, its outburst amplitude is small (about 3~mag), which is significantly lower than typically observed for novae. 
In contrast, although the amplitude is compatible with dwarf novae, the duration of the event is significantly longer than usually observed for this class  (Fig.~\ref{fig:snad283_lc}). Further observations and analysis are therefore required to clarify the origin of this event.

\subsection{Multiple supernovae}

Multiple supernovae refer to cases in which two or more supernovae explode in the same host galaxy (so-called supernova siblings). Such systems are useful tool to investigate possible connections between SN spectroscopic types and their host–galaxy environments~\citep{2025ApJ...981...97S}. In addition, supernova siblings can be used to assess systematic effects in distance measurements, which is important for cosmological applications~\citep{2018A&A...611A..58G}. More generally, they offer a way to probe the properties of the progenitor stellar populations and the host galaxies, including star-formation rate, dust extinction, and metallicity (e.g.~\citealt{2022MNRAS.511..241G}).

In practice, studies of supernova siblings usually focus on cases where multiple transients in the same host galaxy are clearly spatially separated, so that they can be unambiguously identified as independent sources. However, at larger distances the apparent angular size of galaxies decreases, and multiple explosions occurring within a single host may appear unresolved for the particular survey. In such cases, it becomes non-trivial to determine whether we are observing two independent supernovae in the same galaxy, or multiple outbursts from a single source. The latter scenario includes so-called supernova impostors, which are believed to be non-terminal eruptions, often associated with luminous blue variables, such as $\eta$~Carinae. Although these events are also of great astrophysical interest, their physical mechanisms are still not well understood \citep{2011MNRAS.415..773S}.

During the development of the binary classifier and the subsequent anomaly-detection experiments, we identified two galaxies hosting pairs of supernova siblings that, to our knowledge, have not been previously reported in the literature\footnote{We also encountered two additional similar cases in other SNAD-related studies: a re-brightening of the superluminous SN~2018fcg about 3.5 years after the main outburst~\citep{majumder2024}, and two transients, SN~2019rra and AT~2020aewi, occurring in the same host galaxy (Gangler et al., in prep.).}.
Below, we present an initial analysis of the relative positions of these transients within their host galaxies, followed by individual light-curve modelling for each event, performed in the same manner as for the new SN candidates described in Section~\ref{sec:new_sne}.

% These occurrences are scientifically compelling because they provide unique windows into stellar populations, host-galaxy environments and progenitor system properties under nearly identical galactic conditions. Within ZTF’s large and untargeted survey footprint, sibling SNe enable unbiased statistical studies of explosion rates, host metrics (e.g., star-formation rate, dust extinction, metallicity) and relative event frequencies under the same host galaxy environment.

\subsubsection{AT2018mpv/SNAD287 and AT2023kuz/ATLAS23mmv}

\begin{figure*}
\begin{minipage}[h]{1\linewidth}
\center{\includegraphics[width=1\linewidth]{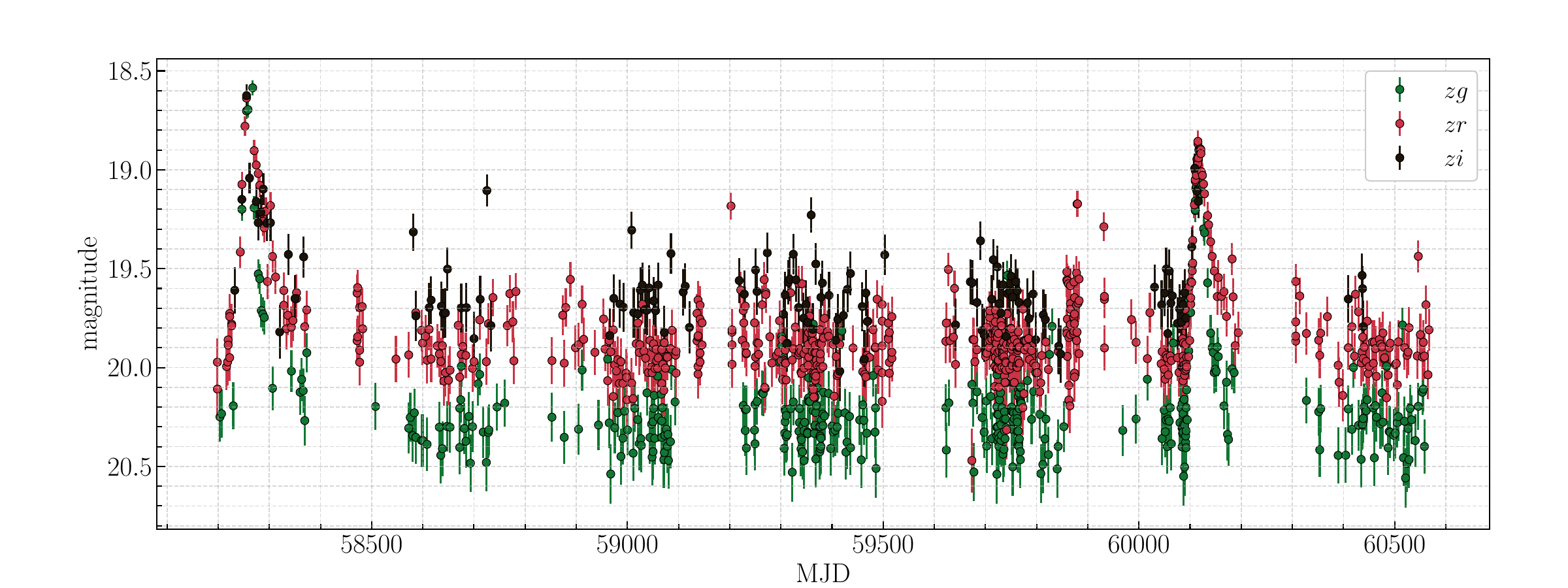}}  \\
\end{minipage}
\vfill
\begin{minipage}[h]{0.47\linewidth}
\center{\includegraphics[width=1\linewidth]{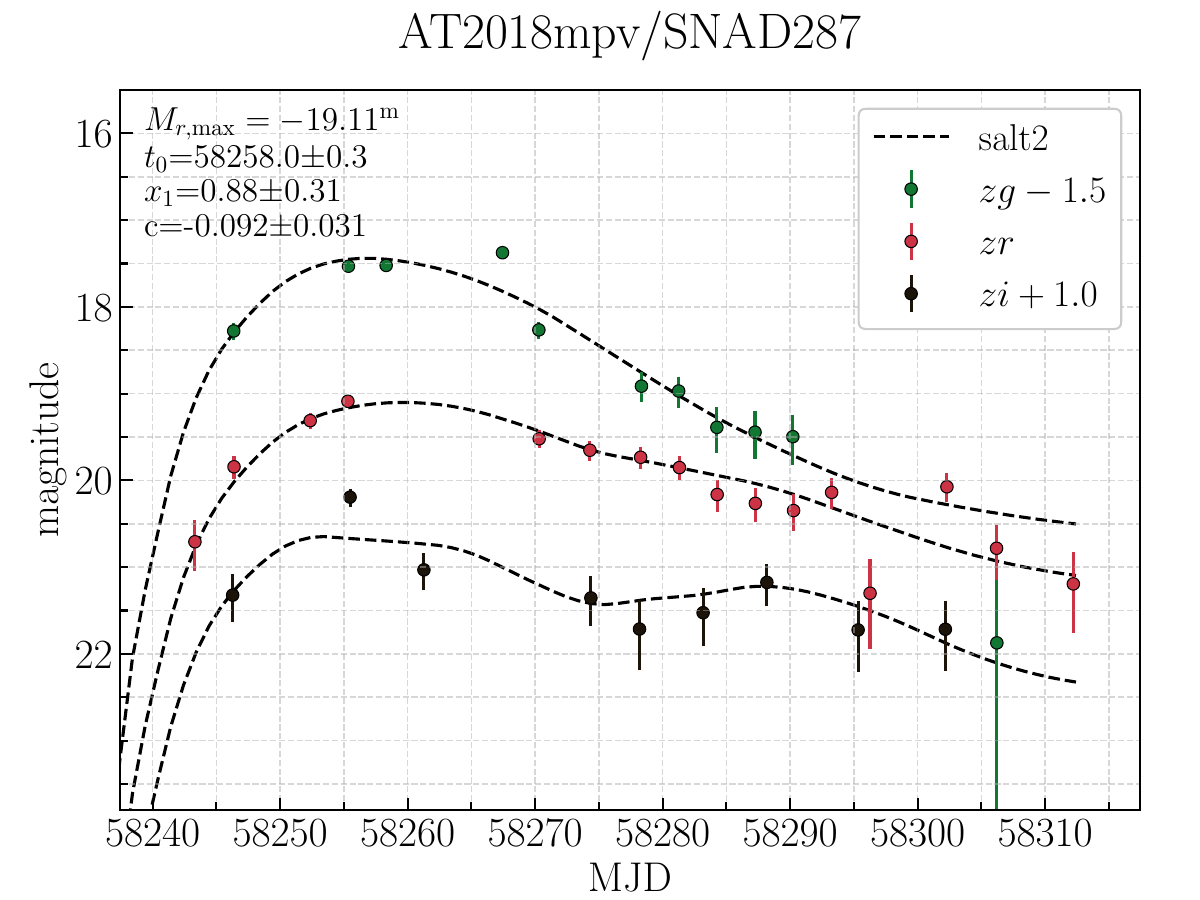}} \\
\end{minipage}
\hfill
\begin{minipage}[h]{0.47\linewidth}
\center{\includegraphics[width=1\linewidth]{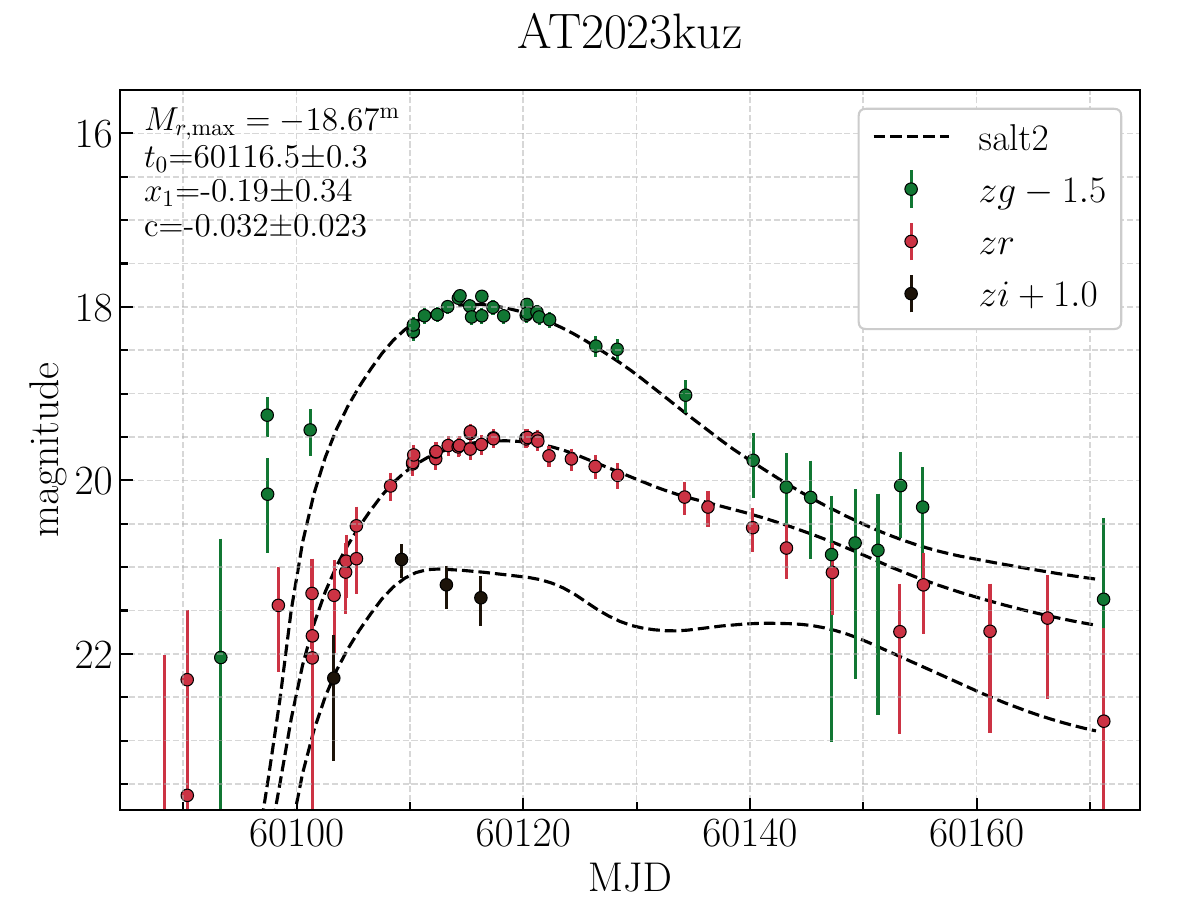}}  \\
\end{minipage}
\caption{Top: ZTF DR23 light curves of the multiple supernovae  AT2018mpv/SNAD287 and AT2023kuz/ATLAS23mmv. 
Bottom: separate \texttt{SNCOSMO} fits to the light curves of both outbursts, with the corresponding best-fit model parameters.}
\label{fig:snad287}
\end{figure*}

The light curve shows two outbursts separated by more than five years (see Fig.~\ref{fig:snad287}). The first outburst appeared in the ZTF alert stream as ZTF19aaltuxk, but was not reported to TNS at the time of the event. We later reported it to TNS as AT2018mpv on 2025-07-30 \citep{2025TNSTR2992....1S}. The second outburst was reported to TNS as AT2023kuz by the ATLAS survey \citep{2023TNSTR1398....1T}.

\begin{figure}
	\centering
    \includegraphics[width=1\linewidth]{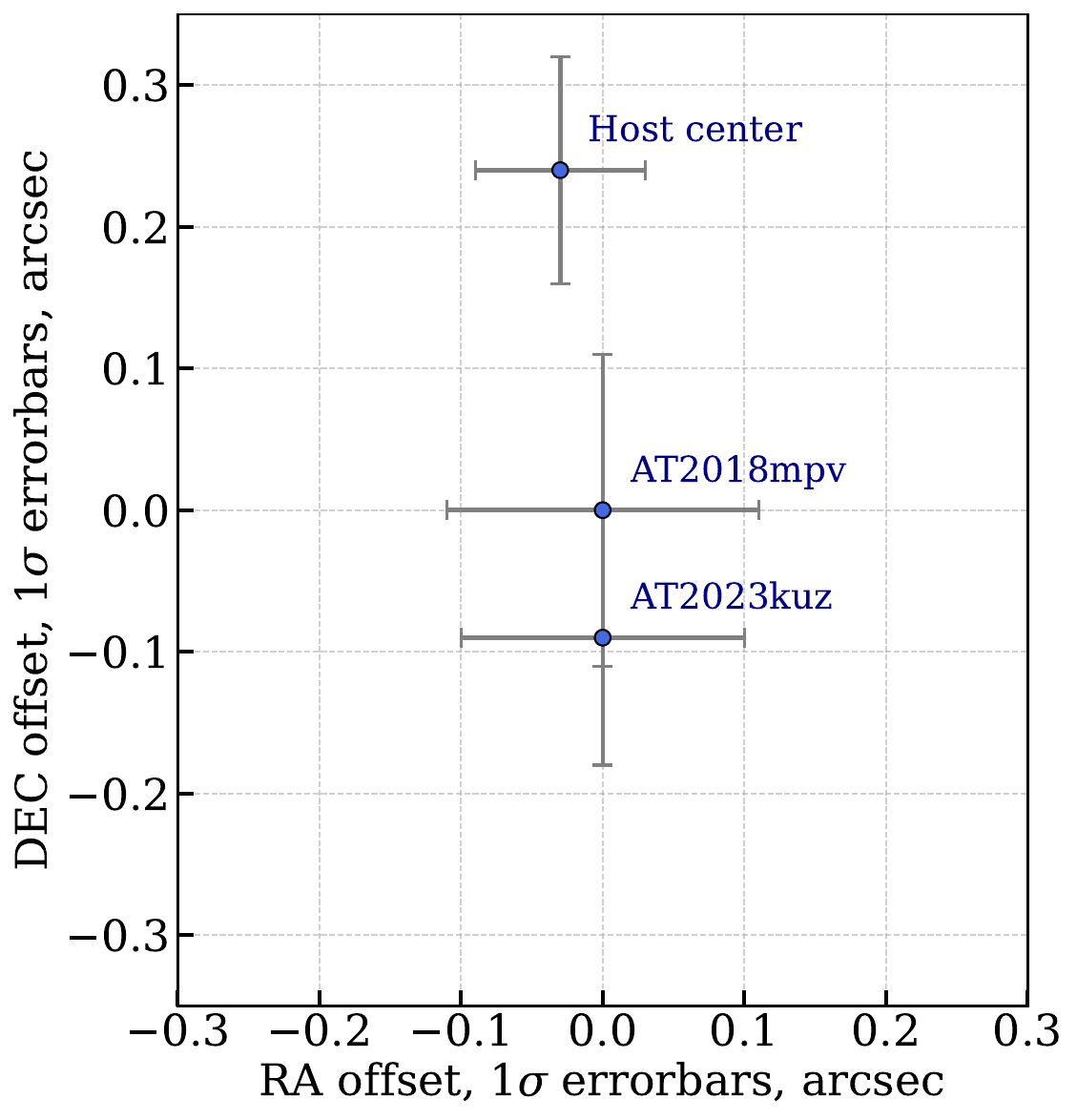}
	\caption{Relative offsets between the two outbursts, AT2018mpv/SNAD287 and AT2023kuz/ATLAS23mmv, and the host galaxy center. The sources positions were measured from stacked ZTF $zr$-band images using two-dimensional Gaussian fitting.}
	\label{fig:533202300011210_offset}
\end{figure}

To measure the offset between the second outburst and the host-galaxy nucleus relative to the first outburst, we used stacked $zr$-band images obtained by ZTF between 2018-05-05 and 2018-07-07 for the first outburst, and between 2023-06-13 and 2023-06-27 for the second outburst. The position of the host galaxy was determined from stacked images obtained between 2022-02-12 and 2022-05-28. Source positions were measured using two-dimensional Gaussian fitting. The measured offset between the two outbursts does not exceed the $1\sigma$ positional uncertainty (see Fig.~\ref{fig:533202300011210_offset}); therefore, we cannot conclude that the two outbursts originate from independent sources.

In order to classify the transients and to determine the redshift of the host galaxy, we obtained a spectrum of the host on 2026 January 29 at 02:52:42~UTC with TDS of the 2.5-m CMO telescope \citep{2020gbar.conf..127S}. The positions of the emission lines yield a redshift of $z=0.09942\pm0.00002$\footnote{\url{https://www.wis-tns.org/object/2018mpv}}.

Using this redshift, we modelled the outbursts with \texttt{SNCOSMO}. The fitting results are shown in the bottom panels of Fig.~\ref{fig:snad287}. For both outbursts, a Type~Ia supernova classification is the most plausible.

\subsubsection{SN2018elp and AT2022mwi}

The light curve shows two outbursts separated by approximately four years. These transients were previously reported to TNS as SN2018elp \citep{2018TNSTR1055....1T} and AT2022mwi.

\begin{figure}
	\centering
    \includegraphics[width=1\linewidth]{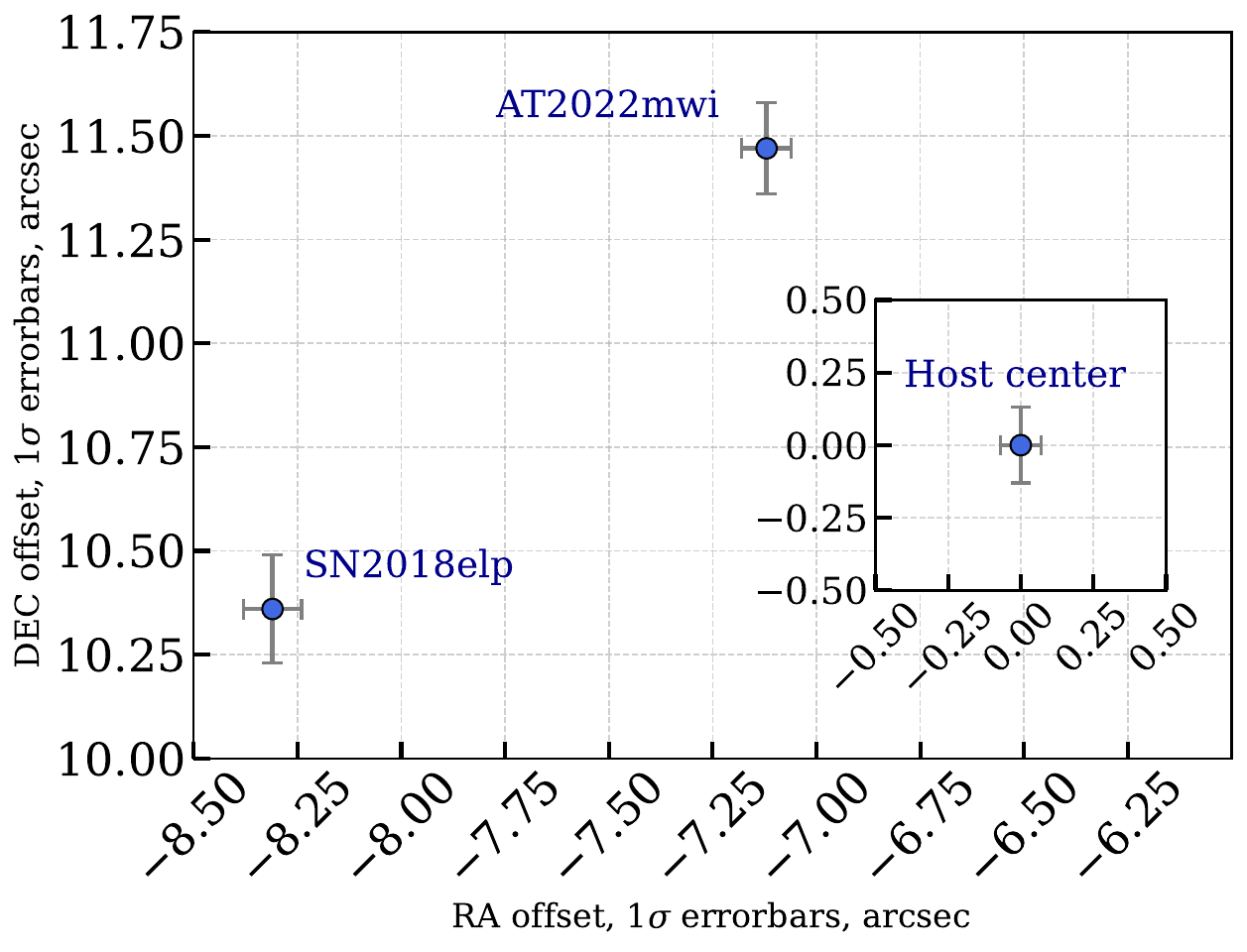}
	\caption{Relative offsets between the two outbursts, SN2018elp and AT2022mwi, and the host galaxy center. The sources positions were measured from stacked ZTF $zr$-band images using two-dimensional Gaussian fitting.}
	\label{fig:580214400011443_offset}
\end{figure}

We measured the positional offsets of both outbursts with respect to the host galaxy (Fig.~\ref{fig:580214400011443_offset}). The position of the first outburst was measured from stacked images obtained between 2018-08-01 and 2018-08-18, and that of the second outburst from stacked images obtained between 2022-06-10 and 2022-06-28. The difference between the two measured positions exceeds $16\sigma$, indicating that these events are spatially distinct. We therefore conclude that they are supernova siblings.

\begin{figure*}
\begin{minipage}[h]{1\linewidth}
\center{\includegraphics[width=1\linewidth]{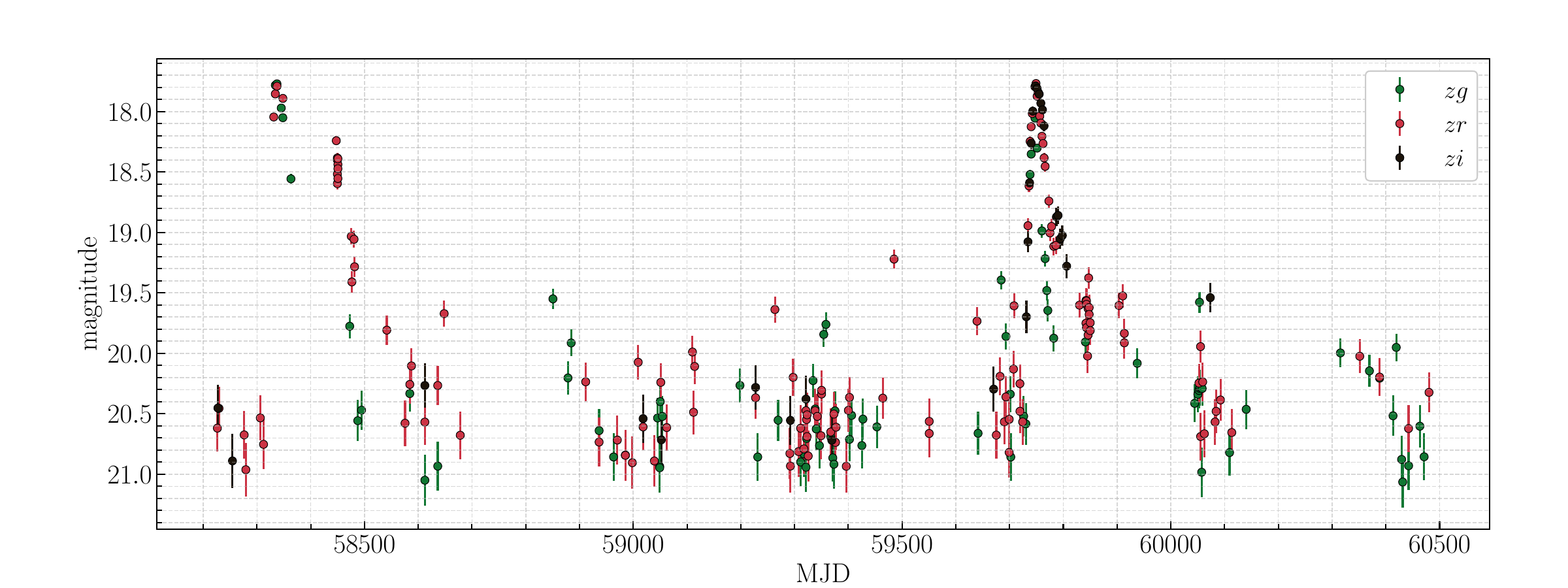}}  \\
\end{minipage}
\vfill
\begin{minipage}[h]{0.47\linewidth}
\center{\includegraphics[width=1\linewidth]{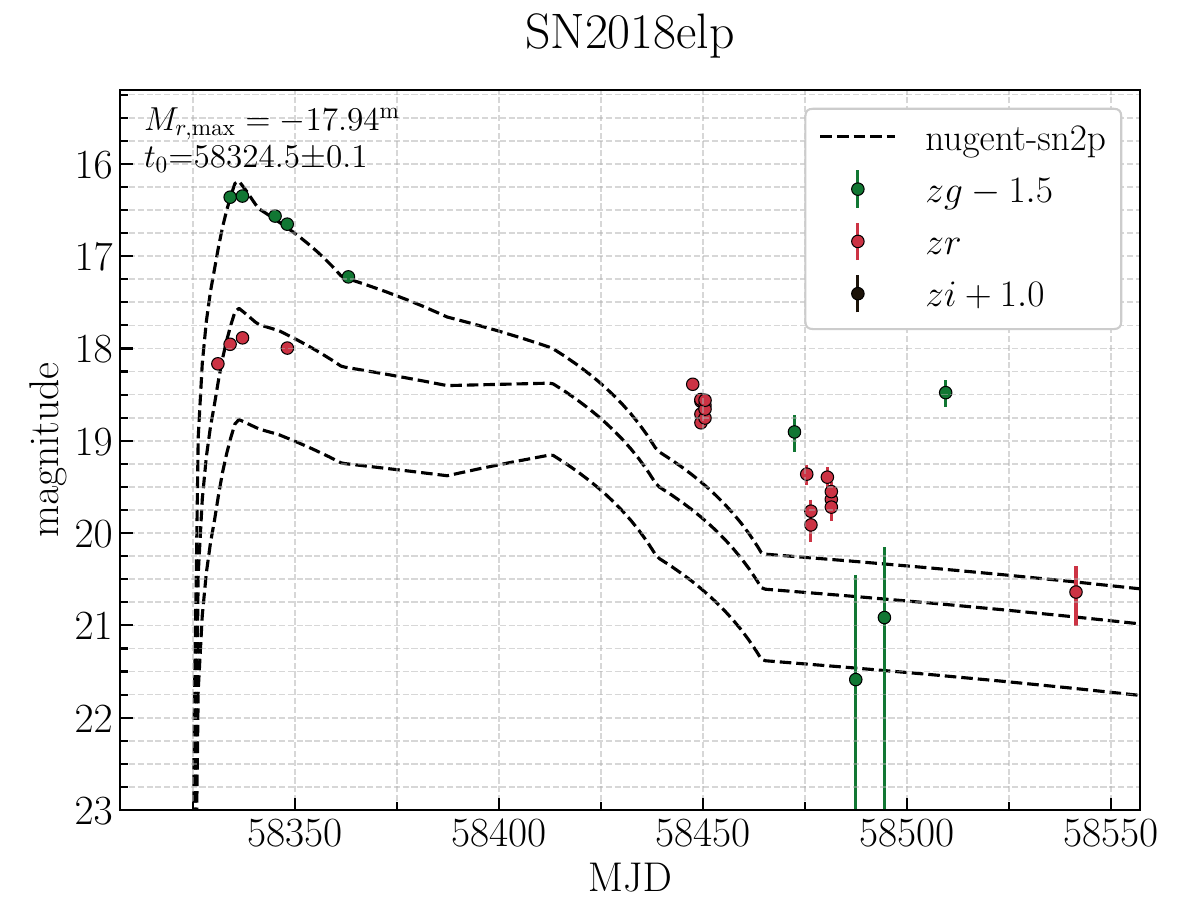}} \\
\end{minipage}
\hfill
\begin{minipage}[h]{0.47\linewidth}
\center{\includegraphics[width=1\linewidth]{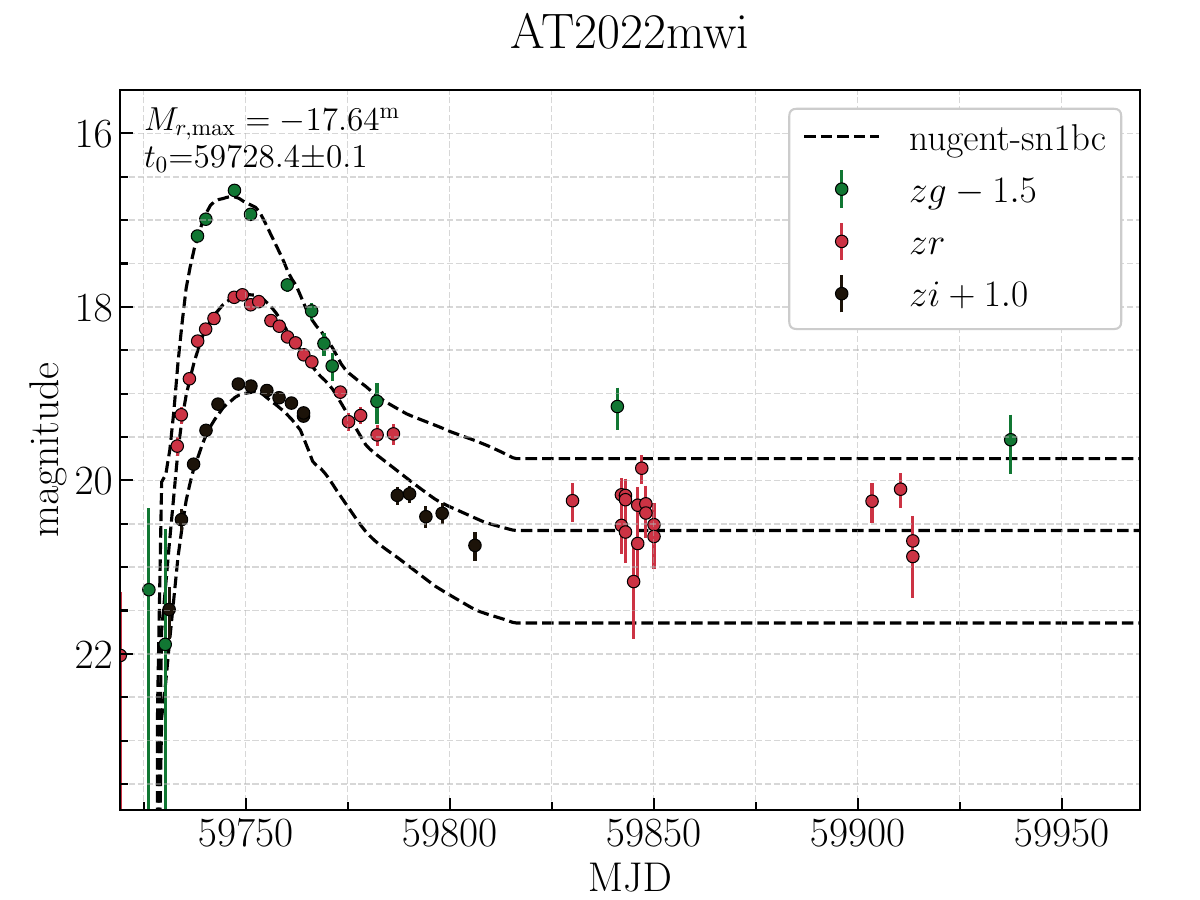}}  \\
\end{minipage}
\caption{Top: ZTF DR23 light curves of the multiple supernovae  SN2018elp and AT2022mwi. 
Bottom: separate \texttt{SNCOSMO} fits to the light curves of both outbursts, with the corresponding best-fit model parameters.}
\label{fig:SN2018el}
\end{figure*}

The first outburst, SN2018elp, is classified in TNS as a Type~II supernova with a measured redshift of $z = 0.03$ \citep{2018TNSCR2128....1S}. We adopted this redshift for the \texttt{SNCOSMO} modelling of both transients. The photometric classification of SN2018elp as SN~IIP is consistent with its spectroscopic classification. For AT2022mwi the best-fitting model corresponds to a Type~Ibc supernova (see Fig.~\ref{fig:SN2018el}).

\section{Conclusions}\label{sec:conclusions}
In this work, we introduced a SN-score designed to enhance active anomaly detection in wide-field time–domain astronomical surveys. Using light curves from ZTF~DR23 and spectroscopically confirmed events from the BTS, we constructed a binary classifier capable of distinguishing SNe from non-SN sources. The classifier achieves high performance (ROC–AUC $\approx 0.98$) and provides a quantitative score that can be incorporated as an additional feature for anomaly detection algorithms.

We integrated this score into the \texttt{PineForest} -- an active anomaly detection framework, and demonstrated that it substantially improves the recovery of SNe when combined with a small number of labeled priors. Our experiments across ten extragalactic ZTF fields show that the augmented feature set enables the algorithm to converge more rapidly toward scientifically relevant candidates, increasing the efficiency of expert-guided exploration of large unlabeled datasets. Importantly, the approach retains the ability to detect diverse astrophysical anomalies rather than simply replicating supervised classification.

Application of the combined methodology resulted in the discovery of seven previously unreported supernova candidates, one AGN candidate, and one unusual Galactic variable star. 
Photometric modelling of the seven supernova candidates with \texttt{SNCOSMO} indicates a sample composed of five SNe~Ia, one SN~IIn, and one SN~IIP. Spectroscopic follow-up of the peculiar source SNAD283 suggests a Galactic helium-rich accreting system, whose long-lasting and low-amplitude outburst is inconsistent with classical nova or dwarf-nova behaviour. 
We also report two host galaxies exhibiting multiple supernova events and present their astrometric and photometric analyses. 
These results demonstrate the ability of the hybrid approach to efficiently discover supernovae and other rare or ambiguous transients in large-scale surveys.

The methods developed here are applicable to ongoing and future time-domain experiments, including the Vera C. Rubin Observatory LSST\footnote{\url{https://www.lsst.org/}}, which will require scalable, adaptive, and expert-steerable systems for identifying and prioritising scientifically interesting objects.

\section*{Acknowledgements}
T.~Semenikhin, M.~Kornilov and M.~Pruzhinskaya  acknowledges support from a Russian Science Foundation grant 24-22-00233, \url{https://rscf.ru/en/project/24-22-00233/} for conceptualization, software development  of the proposed algorithm; conducting experiments and analyzing of obtained results.
T.~Semenikhin acknowledges support  from the Theoretical Physics and Mathematics Advancement Foundation “BASIS” for  formalizing the proposed approach in a repository on GitHub and publishing the data on Zenodo.
Support was provided by Schmidt Sciences, LLC. for K. Malanchev.
The work was carried out using equipment developed with the support of the M.V. Lomonosov Moscow State University Program of Development.

\appendix
\label{app}

\section{List of features}\label{app:features}
Table~\ref{tab:features} lists all features used in this work along with their brief descriptions. A more detailed explanation of each feature can be found in the \texttt{light-curve-feature} Rust crate documentation\footnote{\url{https://docs.rs/light-curve-feature/latest/light\_curve\_feature/features/index.html}}.

{
\renewcommand{\arraystretch}{1.3}
\begin{table*}[h]
\begin{center}
\begin{tabularx}{\textwidth}{
    l
    c
    >{\raggedleft\arraybackslash}X
}
\hline
    Extractor & Number of features & Remark \\
    \hline
    \multicolumn{3}{ c }{Extracted from magnitude} \\
    \hline
     Amplitude & 1 &  \\
     Mean & 1 &      \\
     Kurtosis & 1 &  \\
     Median Absolute Deviation & 1 &  \\
     Anderson Darling Normal & 1 & Unbiased Anderson–Darling normality test statistic \\
     Cusum & 1 &    A range of cumulative sums \\
     BeyondNStd & $1 \times 2$ & Fraction of observations beyond $N\sigma_m$ (for $N = 1,2$) from the mean magnitude \\
     Inter Percentile Range & $1 \times 3$ & Inter-percentile range $Q(1-p) - Q(p)$, where $Q(p)$ is the $p$th quantile of the magnitude distribution (for $p=0.02, 0.1, 0.25$)\\
     Linear Fit & 3 & Slope, its error and reduced $\chi^2$ of the light curve in the linear fit \\
     Linear Trend & 3 & The slope, its error and noise level of the light curve in the linear fit \\
     Periodogram & $2 \times \mathrm{peaks} + 4$ & Peaks of Lomb–Scargle periodogram and periodogram as a meta-feature (for $\mathrm{peaks} = 5$) \\ 
     Reduced $\chi^2$ & 1 &  \\
     Skew & 1 &    Skewness of magnitude $G1$ \\
     Standard Deviation & 1 &    \\
     StetsonK & 1 &   Stetson $K$ coefficient described light curve shape \\
     Weighted Mean & 1 &     \\

     \hline
    \multicolumn{3}{ c }{Extracted from flux} \\
    \hline
    Otsu Split & 4 &  \\
    Mean Variance & 1 & Standard deviation to mean ratio \\
    Bazin Fit & 5 &  \\
    Excess Variance & 1 & Measure of the variability amplitude \\
     \hline
     Total number of features: & 47 & \\
     \hline

\end{tabularx}
\caption{Feature set used in this study.}
\label{tab:features}
\end{center}
\end{table*}
}

\section{Feature importance and class balance}\label{app:selection}
At the initial stage, we trained the binary classifier on a balanced dataset. In this configuration, the sample was relatively small (about 1500 objects), and the resulting performance metrics, including the ROC-AUC, appeared high. However, inspection of the feature-importance diagram revealed that the most influential features were the reduced $\chi^2$ values associated with linear and Bazin function fits, while all other features contributed negligibly. This effect arises because brighter objects have smaller measurement uncertainties. Since the $\chi^2$ statistic is inversely weighted by the observational errors, even small residuals can produce relatively large $\chi^2$ values, despite the model light curve visually providing a good fit to the data points. As a result, the classifier implicitly learned to label faint sources as non-SNe. Since most BTS SNe in our training set are relatively bright, this bias inflated the performance metrics, but led to numerous false positives during inference (most notably bright artifacts). To mitigate this issue, we adopted the sampling strategy described in Section 4.

\bibliographystyle{elsarticle-harv} 
\bibliography{ref}

@ARTICLE{2019PASP..131a8002B,
       author = {{Bellm}, Eric C. and {Kulkarni}, Shrinivas R. and {Graham}, Matthew J. and
         {Dekany}, Richard and {Smith}, Roger M. and {Riddle}, Reed and
         {Masci}, Frank J. and {Helou}, George and {Prince}, Thomas A. and
         {Adams}, Scott M. and {Barbarino}, C. and {Barlow}, Tom and
         {Bauer}, James and {Beck}, Ron and {Belicki}, Justin and
         {Biswas}, Rahul and {Blagorodnova}, Nadejda and {Bodewits}, Dennis and
         {Bolin}, Bryce and {Brinnel}, Valery and {Brooke}, Tim and
         {Bue}, Brian and {Bulla}, Mattia and {Burruss}, Rick and
         {Cenko}, S. Bradley and {Chang}, Chan-Kao and {Connolly}, Andrew and
         {Coughlin}, Michael and {Cromer}, John and {Cunningham}, Virginia and
         {De}, Kishalay and {Delacroix}, Alex and {Desai}, Vandana and
         {Duev}, Dmitry A. and {Eadie}, Gwendolyn and {Farnham}, Tony L. and
         {Feeney}, Michael and {Feindt}, Ulrich and {Flynn}, David and
         {Franckowiak}, Anna and {Frederick}, S. and {Fremling}, C. and
         {Gal-Yam}, Avishay and {Gezari}, Suvi and {Giomi}, Matteo and
         {Goldstein}, Daniel A. and {Golkhou}, V. Zach and {Goobar}, Ariel and
         {Groom}, Steven and {Hacopians}, Eugean and {Hale}, David and
         {Henning}, John and {Ho}, Anna Y.~Q. and {Hover}, David and
         {Howell}, Justin and {Hung}, Tiara and {Huppenkothen}, Daniela and
         {Imel}, David and {Ip}, Wing-Huen and {Ivezi{\'c}}, {\v{Z}}eljko and
         {Jackson}, Edward and {Jones}, Lynne and {Juric}, Mario and
         {Kasliwal}, Mansi M. and {Kaspi}, S. and {Kaye}, Stephen and
         {Kelley}, Michael S.~P. and {Kowalski}, Marek and {Kramer}, Emily and
         {Kupfer}, Thomas and {Landry}, Walter and {Laher}, Russ R. and
         {Lee}, Chien-De and {Lin}, Hsing Wen and {Lin}, Zhong-Yi and
         {Lunnan}, Ragnhild and {Giomi}, Matteo and {Mahabal}, Ashish and
         {Mao}, Peter and {Miller}, Adam A. and {Monkewitz}, Serge and
         {Murphy}, Patrick and {Ngeow}, Chow-Choong and {Nordin}, Jakob and
         {Nugent}, Peter and {Ofek}, Eran and {Patterson}, Maria T. and
         {Penprase}, Bryan and {Porter}, Michael and {Rauch}, Ludwig and
         {Rebbapragada}, Umaa and {Reiley}, Dan and {Rigault}, Mickael and
         {Rodriguez}, Hector and {van Roestel}, Jan and {Rusholme}, Ben and
         {van Santen}, Jakob and {Schulze}, S. and {Shupe}, David L. and
         {Singer}, Leo P. and {Soumagnac}, Maayane T. and {Stein}, Robert and
         {Surace}, Jason and {Sollerman}, Jesper and {Szkody}, Paula and
         {Taddia}, F. and {Terek}, Scott and {Van Sistine}, Angela and
         {van Velzen}, Sjoert and {Vestrand}, W. Thomas and {Walters}, Richard and
         {Ward}, Charlotte and {Ye}, Quan-Zhi and {Yu}, Po-Chieh and {Yan}, Lin and
         {Zolkower}, Jeffry},
        title = "{The Zwicky Transient Facility: System Overview, Performance, and First Results}",
      journal = {\pasp},
         year = 2019,
        month = jan,
       volume = {131},
       number = {995},
        pages = {018002},
          doi = {10.1088/1538-3873/aaecbe},
archivePrefix = {arXiv},
       eprint = {1902.01932},
 primaryClass = {astro-ph.IM},
       adsurl = {https://ui.adsabs.harvard.edu/abs/2019PASP..131a8002B}
}

@ARTICLE{2020AstL...46..836P,
       author = {{Potanin}, S.~A. and {Belinski}, A.~A. and {Dodin}, A.~V. and {Zheltoukhov}, S.~G. and {Lander}, V. Yu. and {Postnov}, K.~A. and {Savvin}, A.~D. and {Tatarnikov}, A.~M. and {Cherepashchuk}, A.~M. and {Cheryasov}, D.~V. and {Chilingarian}, I.~V. and {Shatsky}, N.~I.},
        title = "{Transient Double-Beam Spectrograph for the 2.5-m Telescope of the Caucasus Mountain Observatory of SAI MSU}",
      journal = {Astronomy Letters},
         year = 2020,
        month = dec,
       volume = {46},
       number = {12},
        pages = {836-854},
          doi = {10.1134/S1063773720120038},
archivePrefix = {arXiv},
       eprint = {2011.03061},
 primaryClass = {astro-ph.IM},
       adsurl = {https://ui.adsabs.harvard.edu/abs/2020AstL...46..836P}
}

@ARTICLE{2023PASP..135b4503M,
       author = {{Malanchev}, Konstantin and {Kornilov}, Matwey V. and {Pruzhinskaya}, Maria V. and {Ishida}, Emille E.~O. and {Aleo}, Patrick D. and {Korolev}, Vladimir S. and {Lavrukhina}, Anastasia and {Russeil}, Etienne and {Sreejith}, Sreevarsha and {Volnova}, Alina A. and {Voloshina}, Anastasiya and {Krone-Martins}, Alberto},
        title = "{The SNAD Viewer: Everything You Want to Know about Your Favorite ZTF Object}",
      journal = {\pasp},
         year = 2023,
        month = feb,
       volume = {135},
       number = {1044},
          eid = {024503},
        pages = {024503},
          doi = {10.1088/1538-3873/acb292},
archivePrefix = {arXiv},
       eprint = {2211.07605},
 primaryClass = {astro-ph.IM},
       adsurl = {https://ui.adsabs.harvard.edu/abs/2023PASP..135b4503M}
}

@ARTICLE{2014A&A...568A..22B,
       author = {{Betoule}, M. and {Kessler}, R. and {Guy}, J. and {Mosher}, J. and {Hardin}, D. and {Biswas}, R. and {Astier}, P. and {El-Hage}, P. and {Konig}, M. and {Kuhlmann}, S. and {Marriner}, J. and {Pain}, R. and {Regnault}, N. and {Balland}, C. and {Bassett}, B.~A. and {Brown}, P.~J. and {Campbell}, H. and {Carlberg}, R.~G. and {Cellier-Holzem}, F. and {Cinabro}, D. and {Conley}, A. and {D'Andrea}, C.~B. and {DePoy}, D.~L. and {Doi}, M. and {Ellis}, R.~S. and {Fabbro}, S. and {Filippenko}, A.~V. and {Foley}, R.~J. and {Frieman}, J.~A. and {Fouchez}, D. and {Galbany}, L. and {Goobar}, A. and {Gupta}, R.~R. and {Hill}, G.~J. and {Hlozek}, R. and {Hogan}, C.~J. and {Hook}, I.~M. and {Howell}, D.~A. and {Jha}, S.~W. and {Le Guillou}, L. and {Leloudas}, G. and {Lidman}, C. and {Marshall}, J.~L. and {M{\"o}ller}, A. and {Mour{\~a}o}, A.~M. and {Neveu}, J. and {Nichol}, R. and {Olmstead}, M.~D. and {Palanque-Delabrouille}, N. and {Perlmutter}, S. and {Prieto}, J.~L. and {Pritchet}, C.~J. and {Richmond}, M. and {Riess}, A.~G. and {Ruhlmann-Kleider}, V. and {Sako}, M. and {Schahmaneche}, K. and {Schneider}, D.~P. and {Smith}, M. and {Sollerman}, J. and {Sullivan}, M. and {Walton}, N.~A. and {Wheeler}, C.~J.},
        title = "{Improved cosmological constraints from a joint analysis of the SDSS-II and SNLS supernova samples}",
      journal = {\aap},
         year = 2014,
        month = aug,
       volume = {568},
          eid = {A22},
        pages = {A22},
          doi = {10.1051/0004-6361/201423413},
archivePrefix = {arXiv},
       eprint = {1401.4064},
 primaryClass = {astro-ph.CO},
       adsurl = {https://ui.adsabs.harvard.edu/abs/2014A&A...568A..22B}
}

@ARTICLE{2011MNRAS.415..773S,
       author = {{Smith}, Nathan and {Li}, Weidong and {Silverman}, Jeffrey M. and {Ganeshalingam}, Mohan and {Filippenko}, Alexei V.},
        title = "{Luminous blue variable eruptions and related transients: diversity of progenitors and outburst properties}",
      journal = {\mnras},
         year = 2011,
        month = jul,
       volume = {415},
       number = {1},
        pages = {773-810},
          doi = {10.1111/j.1365-2966.2011.18763.x},
archivePrefix = {arXiv},
       eprint = {1010.3718},
 primaryClass = {astro-ph.SR},
       adsurl = {https://ui.adsabs.harvard.edu/abs/2011MNRAS.415..773S}
}

@ARTICLE{2018A&A...611A..58G,
       author = {{Gall}, C. and {Stritzinger}, M.~D. and {Ashall}, C. and {Baron}, E. and {Burns}, C.~R. and {Hoeflich}, P. and {Hsiao}, E.~Y. and {Mazzali}, P.~A. and {Phillips}, M.~M. and {Filippenko}, A.~V. and {Anderson}, J.~P. and {Benetti}, S. and {Brown}, P.~J. and {Campillay}, A. and {Challis}, P. and {Contreras}, C. and {Elias de la Rosa}, N. and {Folatelli}, G. and {Foley}, R.~J. and {Fraser}, M. and {Holmbo}, S. and {Marion}, G.~H. and {Morrell}, N. and {Pan}, Y.-C. and {Pignata}, G. and {Suntzeff}, N.~B. and {Taddia}, F. and {Torres Robledo}, S. and {Valenti}, S.},
        title = "{Two transitional type Ia supernovae located in the Fornax cluster member NGC 1404: SN 2007on and SN 2011iv}",
      journal = {\aap},
         year = 2018,
        month = mar,
       volume = {611},
          eid = {A58},
        pages = {A58},
          doi = {10.1051/0004-6361/201730886},
archivePrefix = {arXiv},
       eprint = {1707.03823},
 primaryClass = {astro-ph.SR},
       adsurl = {https://ui.adsabs.harvard.edu/abs/2018A&A...611A..58G}
}

@ARTICLE{2025ApJ...981...97S,
       author = {{Salo}, Laura and {Zhou}, Rui and {Johnson}, Samuel and {Kelly}, Patrick and {Jones}, Galin L.},
        title = "{Supernova Siblings and Spectroscopic Host Galaxy Properties}",
      journal = {\apj},
         year = 2025,
        month = mar,
       volume = {981},
       number = {1},
          eid = {97},
        pages = {97},
          doi = {10.3847/1538-4357/adad60},
archivePrefix = {arXiv},
       eprint = {2504.04641},
 primaryClass = {astro-ph.GA},
       adsurl = {https://ui.adsabs.harvard.edu/abs/2025ApJ...981...97S}
}

@ARTICLE{2018TNSTR1055....1T,
       author = {{Tonry}, J. and {Stalder}, B. and {Denneau}, L. and {Heinze}, A. and {Weiland}, H. and {Rest}, A. and {Smith}, K.~W. and {Smartt}, S.~J. and {Young}, D.~R. and {Fulton}, M. and {McBrien}, O. and {O'Neill}, D. and {Clark}, P.},
        title = "{ATLAS Transient Discovery Report for 2018-07-30}",
      journal = {Transient Name Server Discovery Report},
         year = 2018,
        month = jul,
       volume = {2018-1055},
        pages = {1},
       adsurl = {https://ui.adsabs.harvard.edu/abs/2018TNSTR1055....1T}
}

@ARTICLE{2023TNSTR1398....1T,
       author = {{Tonry}, J. and {Denneau}, L. and {Weiland}, H. and {Lawrence}, A. and {Siverd}, R. and {Erasmus}, N. and {Koorts}, W. and {Anderson}, J. and {Jordan}, A. and {Suc}, V. and {Smartt}, S.~J. and {Smith}, K.~W. and {Srivastav}, S. and {Young}, D.~R. and {Nicholl}, M. and {Fulton}, M. and {McCollum}, M. and {Moore}, T. and {Weston}, J. and {Sheng}, X. and {Aamer}, A. and {Shingles}, L. and {Rest}, A. and {Chen}, T.~W. and {Stubbs}, C. and {Sommer}, J. and {Rhodes}, L. and {Andersson}, A.},
        title = "{ATLAS Transient Discovery Report for 2023-06-16}",
      journal = {Transient Name Server Discovery Report},
         year = 2023,
        month = jun,
       volume = {2023-1398},
        pages = {1},
       adsurl = {https://ui.adsabs.harvard.edu/abs/2023TNSTR1398....1T}
}

@ARTICLE{2025TNSTR2992....1S,
       author = {{Semenikhin}, T.},
        title = "{SNAD Transient Discovery Report for 2025-07-30}",
      journal = {Transient Name Server Discovery Report},
         year = 2025,
        month = jul,
       volume = {2025-2992},
        pages = {1},
       adsurl = {https://ui.adsabs.harvard.edu/abs/2025TNSTR2992....1S}
}

@INPROCEEDINGS{2020gbar.conf..127S,
       author = {{Shatsky}, N. and {Belinski}, A. and {Dodin}, A. and {Zheltoukhov}, S. and {Kornilov}, V. and {Postnov}, K. and {Potanin}, S. and {Safonov}, B. and {Tatarnikov}, A. and {Cherepashchuk}, A.},
        title = "{The Caucasian Mountain Observatory of the Sternberg Astronomical Institute: First Six Years of Operation}",
    booktitle = {Ground-Based Astronomy in Russia. 21st Century},
         year = 2020,
       editor = {{Romanyuk}, I.~I. and {Yakunin}, I.~A. and {Valeev}, A.~F. and {Kudryavtsev}, D.~O.},
        month = dec,
        pages = {127-132},
          doi = {10.26119/978-5-6045062-0-2_2020_127},
archivePrefix = {arXiv},
       eprint = {2010.10850},
 primaryClass = {astro-ph.IM},
       adsurl = {https://ui.adsabs.harvard.edu/abs/2020gbar.conf..127S}
}

@article{Perley_2020,
   title={The Zwicky Transient Facility Bright Transient Survey. II. A Public Statistical Sample for Exploring Supernova Demographics*},
   volume={904},
   ISSN={1538-4357},
   url={http://dx.doi.org/10.3847/1538-4357/abbd98},
   DOI={10.3847/1538-4357/abbd98},
   number={1},
   journal={The Astrophysical Journal},
   publisher={American Astronomical Society},
   author={Perley, Daniel A. and Fremling, Christoffer and Sollerman, Jesper and Miller, Adam A. and Dahiwale, Aishwarya S. and Sharma, Yashvi and Bellm, Eric C. and Biswas, Rahul and Brink, Thomas G. and Bruch, Rachel J. and De, Kishalay and Dekany, Richard and Drake, Andrew J. and Duev, Dmitry A. and Filippenko, Alexei V. and Gal-Yam, Avishay and Goobar, Ariel and Graham, Matthew J. and Graham, Melissa L. and Ho, Anna Y. Q. and Irani, Ido and Kasliwal, Mansi M. and Kim, Young-Lo and Kulkarni, S. R. and Mahabal, Ashish and Masci, Frank J. and Modak, Shaunak and Neill, James D. and Nordin, Jakob and Riddle, Reed L. and Soumagnac, Maayane T. and Strotjohann, Nora L. and Schulze, Steve and Taggart, Kirsty and Tzanidakis, Anastasios and Walters, Richard S. and Yan, Lin},
   year={2020},
   month=nov, pages={35} }

@ARTICLE{2021MNRAS.502.5147M,
       author = {{Malanchev}, K.~L. and {Pruzhinskaya}, M.~V. and {Korolev}, V.~S. and {Aleo}, P.~D. and {Kornilov}, M.~V. and {Ishida}, E.~E.~O. and {Krushinsky}, V.~V. and {Mondon}, F. and {Sreejith}, S. and {Volnova}, A.~A. and {Belinski}, A.~A. and {Dodin}, A.~V. and {Tatarnikov}, A.~M. and {Zheltoukhov}, S.~G. and {(The SNAD Team)}},
        title = "{Anomaly detection in the Zwicky Transient Facility DR3}",
      journal = {\mnras},
         year = 2021,
        month = apr,
       volume = {502},
       number = {4},
        pages = {5147-5175},
          doi = {10.1093/mnras/stab316},
archivePrefix = {arXiv},
       eprint = {2012.01419},
 primaryClass = {astro-ph.IM},
       adsurl = {https://ui.adsabs.harvard.edu/abs/2021MNRAS.502.5147M}
}

@article{SEMENIKHIN2025100919,
title = {Real-bogus scores for active anomaly detection},
journal = {Astronomy and Computing},
volume = {51},
pages = {100919},
year = {2025},
issn = {2213-1337},
doi = {https://doi.org/10.1016/j.ascom.2024.100919},
url = {https://www.sciencedirect.com/science/article/pii/S2213133724001343},
author = {T.A. Semenikhin and M.V. Kornilov and M.V. Pruzhinskaya and A.D. Lavrukhina and E. Russeil and E. Gangler and E.E.O. Ishida and V.S. Korolev and K.L. Malanchev and A.A. Volnova and S. Sreejith}
}

@article{KORNILOV2025100960,
title = {Coniferest: A complete active anomaly detection framework},
journal = {Astronomy and Computing},
volume = {52},
pages = {100960},
year = {2025},
issn = {2213-1337},
doi = {https://doi.org/10.1016/j.ascom.2025.100960},
url = {https://www.sciencedirect.com/science/article/pii/S2213133725000332},
author = {M.V. Kornilov and V.S. Korolev and K.L. Malanchev and A.D. Lavrukhina and E. Russeil and T.A. Semenikhin and E. Gangler and E.E.O. Ishida and M.V. Pruzhinskaya and A.A. Volnova and S. Sreejith}
}

@INPROCEEDINGS{isolationforest,
  author={Liu, Fei Tony and Ting, Kai Ming and Zhou, Zhi-Hua},
  booktitle={2008 Eighth IEEE International Conference on Data Mining}, 
  title={Isolation Forest}, 
  year={2008},
  volume={},
  number={},
  pages={413-422},
  doi={10.1109/ICDM.2008.17}}

@misc{2009arXiv0912.0201L,
       author = {{LSST Science Collaboration} and {Abell}, Paul A. and {Allison}, Julius and {Anderson}, Scott F. and {Andrew}, John R. and {Angel}, J. Roger P. and {Armus}, Lee and {Arnett}, David and {Asztalos}, S.~J. and {Axelrod}, Tim S. and {Bailey}, Stephen and {Ballantyne}, D.~R. and {Bankert}, Justin R. and {Barkhouse}, Wayne A. and {Barr}, Jeffrey D. and {Barrientos}, L. Felipe and {Barth}, Aaron J. and {Bartlett}, James G. and {Becker}, Andrew C. and {Becla}, Jacek and {Beers}, Timothy C. and {Bernstein}, Joseph P. and {Biswas}, Rahul and {Blanton}, Michael R. and {Bloom}, Joshua S. and {Bochanski}, John J. and {Boeshaar}, Pat and {Borne}, Kirk D. and {Bradac}, Marusa and {Brandt}, W.~N. and {Bridge}, Carrie R. and {Brown}, Michael E. and {Brunner}, Robert J. and {Bullock}, James S. and {Burgasser}, Adam J. and {Burge}, James H. and {Burke}, David L. and {Cargile}, Phillip A. and {Chandrasekharan}, Srinivasan and {Chartas}, George and {Chesley}, Steven R. and {Chu}, You-Hua and {Cinabro}, David and {Claire}, Mark W. and {Claver}, Charles F. and {Clowe}, Douglas and {Connolly}, A.~J. and {Cook}, Kem H. and {Cooke}, Jeff and {Cooray}, Asantha and {Covey}, Kevin R. and {Culliton}, Christopher S. and {de Jong}, Roelof and {de Vries}, Willem H. and {Debattista}, Victor P. and {Delgado}, Francisco and {Dell'Antonio}, Ian P. and {Dhital}, Saurav and {Di Stefano}, Rosanne and {Dickinson}, Mark and {Dilday}, Benjamin and {Djorgovski}, S.~G. and {Dobler}, Gregory and {Donalek}, Ciro and {Dubois-Felsmann}, Gregory and {Durech}, Josef and {Eliasdottir}, Ardis and {Eracleous}, Michael and {Eyer}, Laurent and {Falco}, Emilio E. and {Fan}, Xiaohui and {Fassnacht}, Christopher D. and {Ferguson}, Harry C. and {Fernandez}, Yanga R. and {Fields}, Brian D. and {Finkbeiner}, Douglas and {Figueroa}, Eduardo E. and {Fox}, Derek B. and {Francke}, Harold and {Frank}, James S. and {Frieman}, Josh and {Fromenteau}, Sebastien and {Furqan}, Muhammad and {Galaz}, Gaspar and {Gal-Yam}, A. and {Garnavich}, Peter and {Gawiser}, Eric and {Geary}, John and {Gee}, Perry and {Gibson}, Robert R. and {Gilmore}, Kirk and {Grace}, Emily A. and {Green}, Richard F. and {Gressler}, William J. and {Grillmair}, Carl J. and {Habib}, Salman and {Haggerty}, J.~S. and {Hamuy}, Mario and {Harris}, Alan W. and {Hawley}, Suzanne L. and {Heavens}, Alan F. and {Hebb}, Leslie and {Henry}, Todd J. and {Hileman}, Edward and {Hilton}, Eric J. and {Hoadley}, Keri and {Holberg}, J.~B. and {Holman}, Matt J. and {Howell}, Steve B. and {Infante}, Leopoldo and {Ivezic}, Zeljko and {Jacoby}, Suzanne H. and {Jain}, Bhuvnesh and {R} and {Jedicke} and {Jee}, M. James and {Garrett Jernigan}, J. and {Jha}, Saurabh W. and {Johnston}, Kathryn V. and {Jones}, R. Lynne and {Juric}, Mario and {Kaasalainen}, Mikko and {Styliani} and {Kafka} and {Kahn}, Steven M. and {Kaib}, Nathan A. and {Kalirai}, Jason and {Kantor}, Jeff and {Kasliwal}, Mansi M. and {Keeton}, Charles R. and {Kessler}, Richard and {Knezevic}, Zoran and {Kowalski}, Adam and {Krabbendam}, Victor L. and {Krughoff}, K. Simon and {Kulkarni}, Shrinivas and {Kuhlman}, Stephen and {Lacy}, Mark and {Lepine}, Sebastien and {Liang}, Ming and {Lien}, Amy and {Lira}, Paulina and {Long}, Knox S. and {Lorenz}, Suzanne and {Lotz}, Jennifer M. and {Lupton}, R.~H. and {Lutz}, Julie and {Macri}, Lucas M. and {Mahabal}, Ashish A. and {Mandelbaum}, Rachel and {Marshall}, Phil and {May}, Morgan and {McGehee}, Peregrine M. and {Meadows}, Brian T. and {Meert}, Alan and {Milani}, Andrea and {Miller}, Christopher J. and {Miller}, Michelle and {Mills}, David and {Minniti}, Dante and {Monet}, David and {Mukadam}, Anjum S. and {Nakar}, Ehud and {Neill}, Douglas R. and {Newman}, Jeffrey A. and {Nikolaev}, Sergei and {Nordby}, Martin and {O'Connor}, Paul and {Oguri}, Masamune and {Oliver}, John and {Olivier}, Scot S. and {Olsen}, Julia K. and {Olsen}, Knut and {Olszewski}, Edward W. and {Oluseyi}, Hakeem and {Padilla}, Nelson D. and {Parker}, Alex and {Pepper}, Joshua and {Peterson}, John R. and {Petry}, Catherine and {Pinto}, Philip A. and {Pizagno}, James L. and {Popescu}, Bogdan and {Prsa}, Andrej and {Radcka}, Veljko and {Raddick}, M. Jordan and {Rasmussen}, Andrew and {Rau}, Arne and {Rho}, Jeonghee and {Rhoads}, James E. and {Richards}, Gordon T. and {Ridgway}, Stephen T. and {Robertson}, Brant E. and {Roskar}, Rok and {Saha}, Abhijit and {Sarajedini}, Ata and {Scannapieco}, Evan and {Schalk}, Terry and {Schindler}, Rafe and {Schmidt}, Samuel and {Schmidt}, Sarah and {Schneider}, Donald P. and {Schumacher}, German and {Scranton}, Ryan and {Sebag}, Jacques and {Seppala}, Lynn G. and {Shemmer}, Ohad and {Simon}, Joshua D. and {Sivertz}, M. and {Smith}, Howard A. and {Allyn Smith}, J. and {Smith}, Nathan and {Spitz}, Anna H. and {Stanford}, Adam and {Stassun}, Keivan G. and {Strader}, Jay and {Strauss}, Michael A. and {Stubbs}, Christopher W. and {Sweeney}, Donald W. and {Szalay}, Alex and {Szkody}, Paula and {Takada}, Masahiro and {Thorman}, Paul and {Trilling}, David E. and {Trimble}, Virginia and {Tyson}, Anthony and {Van Berg}, Richard and {Vanden Berk}, Daniel and {VanderPlas}, Jake and {Verde}, Licia and {Vrsnak}, Bojan and {Walkowicz}, Lucianne M. and {Wandelt}, Benjamin D. and {Wang}, Sheng and {Wang}, Yun and {Warner}, Michael and {Wechsler}, Risa H. and {West}, Andrew A. and {Wiecha}, Oliver and {Williams}, Benjamin F. and {Willman}, Beth and {Wittman}, David and {Wolff}, Sidney C. and {Wood-Vasey}, W. Michael and {Wozniak}, Przemek and {Young}, Patrick and {Zentner}, Andrew and {Zhan}, Hu},
        title = "{LSST Science Book, Version 2.0}",
      journal = {arXiv e-prints},
         year = 2009,
        month = dec,
          eid = {arXiv:0912.0201},
        pages = {arXiv:0912.0201},
          doi = {10.48550/arXiv.0912.0201},
archivePrefix = {arXiv},
       eprint = {0912.0201},
 primaryClass = {astro-ph.IM},
       adsurl = {https://ui.adsabs.harvard.edu/abs/2009arXiv0912.0201L}
}

@article{rand,
author = {Breiman, Leo},
year = {2001},
title = {Random Forests},
journal = {Machine Learning},
pages = {5-32},
volume = {45},
issue = {1},
url = {https://doi.org/10.1023/A:1010933404324},
doi = {10.1023/A:1010933404324},
}

@article{Bloom_2012,
   title={Automating Discovery and Classification of Transients and Variable Stars in the Synoptic Survey Era},
   volume={124},
   ISSN={1538-3873},
   url={http://dx.doi.org/10.1086/668468},
   DOI={10.1086/668468},
   number={921},
   journal={Publications of the Astronomical Society of the Pacific},
   publisher={IOP Publishing},
   author={Bloom, J. S. and Richards, J. W. and Nugent, P. E. and Quimby, R. M. and Kasliwal, M. M. and Starr, D. L. and Poznanski, D. and Ofek, E. O. and Cenko, S. B. and Butler, N. R. and Kulkarni, S. R. and Gal-Yam, A. and Law, N.},
   year={2012},
   month=nov, pages={1175–1196} }

@article{Wright_2015,
    author = {Wright, D. E. and Smartt, S. J. and Smith, K. W. and Miller, P. and Kotak, R. and Rest, A. and Burgett, W. S. and Chambers, K. C. and Flewelling, H. and Hodapp, K. W. and Huber, M. and Jedicke, R. and Kaiser, N. and Metcalfe, N. and Price, P. A. and Tonry, J. L. and Wainscoat, R. J. and Waters, C.},
    title = {Machine learning for transient discovery in Pan-STARRS1 difference imaging},
    journal = {Monthly Notices of the Royal Astronomical Society},
    volume = {449},
    number = {1},
    pages = {451-466},
    year = {2015},
    month = {03},
    issn = {0035-8711},
    doi = {10.1093/mnras/stv292},
    url = {https://doi.org/10.1093/mnras/stv292},
    eprint = {https://academic.oup.com/mnras/article-pdf/449/1/451/4139177/stv292.pdf},
}

@article{Richards_2011,
doi = {10.1088/0004-637X/733/1/10},
url = {https://doi.org/10.1088/0004-637X/733/1/10},
year = {2011},
month = {apr},
publisher = {The American Astronomical Society},
volume = {733},
number = {1},
pages = {10},
author = {Richards, Joseph W. and Starr, Dan L. and Butler, Nathaniel R. and Bloom, Joshua S. and Brewer, John M. and Crellin-Quick, Arien and Higgins, Justin and Kennedy, Rachel and Rischard, Maxime},
title = {ON MACHINE-LEARNED CLASSIFICATION OF VARIABLE STARS WITH SPARSE AND NOISY TIME-SERIES DATA},
journal = {The Astrophysical Journal}
}

@article{Duev_2019,
   title={Real-bogus classification for the Zwicky Transient Facility using deep learning},
   volume={489},
   ISSN={1365-2966},
   url={http://dx.doi.org/10.1093/mnras/stz2357},
   DOI={10.1093/mnras/stz2357},
   number={3},
   journal={Monthly Notices of the Royal Astronomical Society},
   publisher={Oxford University Press (OUP)},
   author={Duev, Dmitry A and Mahabal, Ashish and Masci, Frank J and Graham, Matthew J and Rusholme, Ben and Walters, Richard and Karmarkar, Ishani and Frederick, Sara and Kasliwal, Mansi M and Rebbapragada, Umaa and Ward, Charlotte},
   year={2019},
   month=aug, pages={3582–3590} }

@INPROCEEDINGS{2002SPIE.4836...61A,
       author = {{Aldering}, Greg and {Adam}, G. and {Antilogus}, P. and {Astier}, Pierre and {Bacon}, R. and {Bongard}, S. and {Bonnaud}, C. and {Copin}, Y. and {Hardin}, Delphine and {Henault}, Francois and {Howell}, Dale A. and {Lemonnier}, Jean-Pierre and {Levy}, Jean-Michel and {Loken}, Stewart C. and {Nugent}, Peter E. and {Pain}, Reynald and {Pecontal}, A. and {Pecontal}, E. and {Perlmutter}, Saul and {Quimby}, Robert M. and {Schahmaneche}, K. and {Smadja}, G. and {Wood-Vasey}, W. Michael},
        title = "{Overview of the Nearby Supernova Factory}",
    booktitle = {Survey and Other Telescope Technologies and Discoveries},
         year = 2002,
       editor = {{Tyson}, J. Anthony and {Wolff}, Sidney},
       series = {Society of Photo-Optical Instrumentation Engineers (SPIE) Conference Series},
       volume = {4836},
        month = dec,
        pages = {61-72},
          doi = {10.1117/12.458107},
       adsurl = {https://ui.adsabs.harvard.edu/abs/2002SPIE.4836...61A}
}

@ARTICLE{2007ApJ...665.1246B,
       author = {{Bailey}, S. and {Aragon}, C. and {Romano}, R. and {Thomas}, R.~C. and {Weaver}, B.~A. and {Wong}, D.},
        title = "{How to Find More Supernovae with Less Work: Object Classification Techniques for Difference Imaging}",
      journal = {\apj},
         year = 2007,
        month = aug,
       volume = {665},
       number = {2},
        pages = {1246-1253},
          doi = {10.1086/519832},
archivePrefix = {arXiv},
       eprint = {0705.0493},
 primaryClass = {astro-ph},
       adsurl = {https://ui.adsabs.harvard.edu/abs/2007ApJ...665.1246B}
}

@ARTICLE{2022MNRAS.511..241G,
       author = {{Graham}, Melissa L. and {Fremling}, Christoffer and {Perley}, Daniel A. and {Biswas}, Rahul and {Phillips}, Christopher A. and {Sollerman}, Jesper and {Nugent}, Peter E. and {Nance}, Sarafina and {Dhawan}, Suhail and {Nordin}, Jakob and {Goobar}, Ariel and {Miller}, Adam and {Neill}, James D. and {Hall}, Xander J. and {Hankins}, Matthew J. and {Duev}, Dmitry A. and {Kasliwal}, Mansi M. and {Rigault}, Mickael and {Bellm}, Eric C. and {Hale}, David and {Mr{\'o}z}, Przemek and {Kulkarni}, S.~R.},
        title = "{Supernova siblings and their parent galaxies in the Zwicky Transient Facility Bright Transient Survey}",
      journal = {\mnras},
         year = 2022,
        month = mar,
       volume = {511},
       number = {1},
        pages = {241-254},
          doi = {10.1093/mnras/stab3802},
archivePrefix = {arXiv},
       eprint = {2112.14819},
 primaryClass = {astro-ph.HE},
       adsurl = {https://ui.adsabs.harvard.edu/abs/2022MNRAS.511..241G}
}

@misc{majumder2024,
      title={Superluminous supernova search with PineForest}, 
      author={T. Majumder and M. V. Pruzhinskaya and E. E. O. Ishida and K. L. Malanchev and T. A. Semenikhin},
      year={2024},
      eprint={2410.21077},
      archivePrefix={arXiv},
      primaryClass={astro-ph.IM},
      url={https://arxiv.org/abs/2410.21077}, 
}

@ARTICLE{2011ApJ...737..103S,
       author = {{Schlafly}, Edward F. and {Finkbeiner}, Douglas P.},
        title = "{Measuring Reddening with Sloan Digital Sky Survey Stellar Spectra and Recalibrating SFD}",
      journal = {\apj},
         year = 2011,
        month = aug,
       volume = {737},
       number = {2},
          eid = {103},
        pages = {103},
          doi = {10.1088/0004-637X/737/2/103},
archivePrefix = {arXiv},
       eprint = {1012.4804},
 primaryClass = {astro-ph.GA},
       adsurl = {https://ui.adsabs.harvard.edu/abs/2011ApJ...737..103S}
}

@ARTICLE{2018TNSCR2128....1S,
       author = {{Stein}, R. and {Callis}, E. and {Kostrzewa-Rutkowska}, Z. and {Fraser}, M. and {Yaron}, O.},
        title = "{ePESSTO Transient Classification Report for 2018-08-13}",
      journal = {Transient Name Server Classification Report},
         year = 2018,
        month = aug,
       volume = {2018-2128},
        pages = {1},
       adsurl = {https://ui.adsabs.harvard.edu/abs/2018TNSCR2128....1S}
}

@ARTICLE{2025arXiv250503509G,
       author = {{G{\'o}mez}, Pablo and {Ruhberg}, Laslo E. and {Nardone}, Maria Teresa and {O'Ryan}, David},
        title = "{AnomalyMatch: Discovering Rare Objects of Interest with Semi-supervised and Active Learning}",
      journal = {arXiv e-prints},
         year = 2025,
        month = may,
          eid = {arXiv:2505.03509},
        pages = {arXiv:2505.03509},
          doi = {10.48550/arXiv.2505.03509},
archivePrefix = {arXiv},
       eprint = {2505.03509},
 primaryClass = {stat.ML},
       adsurl = {https://ui.adsabs.harvard.edu/abs/2025arXiv250503509G}
}

@article{pruzh2023,
	author = {{Pruzhinskaya, M. V.} and {Ishida, E. E. O.} and {Novinskaya, A. K.} and {Russeil, E.} and {Volnova, A. A.} and {Malanchev, K. L.} and {Kornilov, M. V.} and {Aleo, P. D.} and {Korolev, V. S.} and {Krushinsky, V. V.} and {Sreejith, S.} and {Gangler, E.}},
	title = {Supernova search with active learning in ZTF DR3},
	DOI= "10.1051/0004-6361/202245172",
	url= "https://doi.org/10.1051/0004-6361/202245172",
	journal = {A\&A},
	year = 2023,
	volume = 672,
	pages = "A111",
}

@article{Mahabal_2019,
doi = {10.1088/1538-3873/aaf3fa},
url = {https://doi.org/10.1088/1538-3873/aaf3fa},
year = {2019},
month = {jan},
publisher = {The Astronomical Society of the Pacific},
volume = {131},
number = {997},
pages = {038002},
author = {Mahabal, Ashish and Rebbapragada, Umaa and Walters, Richard and Masci, Frank J. and Blagorodnova, Nadejda and Roestel, Jan van and Ye, Quan-Zhi and Biswas, Rahul and Burdge, Kevin and Chang, Chan-Kao and Duev, Dmitry A. and Golkhou, V. Zach and Miller, Adam A. and Nordin, Jakob and Ward, Charlotte and Adams, Scott and Bellm, Eric C. and Branton, Doug and Bue, Brian and Cannella, Chris and Connolly, Andrew and Dekany, Richard and Feindt, Ulrich and Hung, Tiara and Fortson, Lucy and Frederick, Sara and Fremling, C. and Gezari, Suvi and Graham, Matthew and Groom, Steven and Kasliwal, Mansi M. and Kulkarni, Shrinivas and Kupfer, Thomas and Lin, Hsing Wen and Lintott, Chris and Lunnan, Ragnhild and Parejko, John and Prince, Thomas A. and Riddle, Reed and Rusholme, Ben and Saunders, Nicholas and Sedaghat, Nima and Shupe, David L. and Singer, Leo P. and Soumagnac, Maayane T. and Szkody, Paula and Tachibana, Yutaro and Tirumala, Kushal and Velzen, Sjoert van and Wright, Darryl},
title = {Machine Learning for the Zwicky Transient Facility},
journal = {Publications of the Astronomical Society of the Pacific}
}

@ARTICLE{2019PASP..131a8003M,
       author = {{Masci}, Frank J. and {Laher}, Russ R. and {Rusholme}, Ben and {Shupe}, David L. and {Groom}, Steven and {Surace}, Jason and {Jackson}, Edward and {Monkewitz}, Serge and {Beck}, Ron and {Flynn}, David and {Terek}, Scott and {Landry}, Walter and {Hacopians}, Eugean and {Desai}, Vandana and {Howell}, Justin and {Brooke}, Tim and {Imel}, David and {Wachter}, Stefanie and {Ye}, Quan-Zhi and {Lin}, Hsing-Wen and {Cenko}, S. Bradley and {Cunningham}, Virginia and {Rebbapragada}, Umaa and {Bue}, Brian and {Miller}, Adam A. and {Mahabal}, Ashish and {Bellm}, Eric C. and {Patterson}, Maria T. and {Juri{\'c}}, Mario and {Golkhou}, V. Zach and {Ofek}, Eran O. and {Walters}, Richard and {Graham}, Matthew and {Kasliwal}, Mansi M. and {Dekany}, Richard G. and {Kupfer}, Thomas and {Burdge}, Kevin and {Cannella}, Christopher B. and {Barlow}, Tom and {Van Sistine}, Angela and {Giomi}, Matteo and {Fremling}, Christoffer and {Blagorodnova}, Nadejda and {Levitan}, David and {Riddle}, Reed and {Smith}, Roger M. and {Helou}, George and {Prince}, Thomas A. and {Kulkarni}, Shrinivas R.},
        title = "{The Zwicky Transient Facility: Data Processing, Products, and Archive}",
      journal = {\pasp},
         year = 2019,
        month = jan,
       volume = {131},
       number = {995},
        pages = {018003},
          doi = {10.1088/1538-3873/aae8ac},
archivePrefix = {arXiv},
       eprint = {1902.01872},
 primaryClass = {astro-ph.IM},
       adsurl = {https://ui.adsabs.harvard.edu/abs/2019PASP..131a8003M}
}

@ARTICLE{karpov2023,
       author = {{Karpov}, S. and {Peloton}, J.},
        title = "{The rate of satellite glints in ZTF and LSST sky surveys}",
      journal = {Contributions of the Astronomical Observatory Skalnate Pleso},
         year = 2023,
        month = dec,
       volume = {53},
       number = {4},
        pages = {69-80},
          doi = {10.31577/caosp.2023.53.4.69},
archivePrefix = {arXiv},
       eprint = {2310.17322},
 primaryClass = {astro-ph.IM},
       adsurl = {https://ui.adsabs.harvard.edu/abs/2023CoSka..53d..69K}
}

@article{ishida2021,
	author = {{Ishida, E. E. O.} and {Kornilov, M. V.} and {Malanchev, K. L.} and {Pruzhinskaya, M. V.} and {Volnova, A. A.} and {Korolev, V. S.} and {Mondon, F.} and {Sreejith, S.} and {Malancheva, A. A.} and {Das, S.}},
	title = {Active anomaly detection for time-domain discoveries},
	DOI= "10.1051/0004-6361/202037709",
	url= "https://doi.org/10.1051/0004-6361/202037709",
	journal = {A\&A},
	year = 2021,
	volume = 650,
	pages = "A195",
}

@article{webb2020,
    author = {Webb, Sara and Lochner, Michelle and Muthukrishna, Daniel and Cooke, Jeff and Flynn, Chris and Mahabal, Ashish and Goode, Simon and Andreoni, Igor and Pritchard, Tyler and Abbott, Timothy M C},
    title = {Unsupervised machine learning for transient discovery in deeper, wider, faster light curves},
    journal = {Monthly Notices of the Royal Astronomical Society},
    volume = {498},
    number = {3},
    pages = {3077-3094},
    year = {2020},
    month = {09},
    issn = {0035-8711},
    doi = {10.1093/mnras/staa2395},
    url = {https://doi.org/10.1093/mnras/staa2395},
    eprint = {https://academic.oup.com/mnras/article-pdf/498/3/3077/33780541/staa2395.pdf},
}

@ARTICLE{2019MNRAS.489.3591P,
       author = {{Pruzhinskaya}, M.~V. and {Malanchev}, K.~L. and {Kornilov}, M.~V. and {Ishida}, E.~E.~O. and {Mondon}, F. and {Volnova}, A.~A. and {Korolev}, V.~S.},
        title = "{Anomaly detection in the Open Supernova Catalog}",
      journal = {\mnras},
         year = 2019,
        month = nov,
       volume = {489},
       number = {3},
        pages = {3591-3608},
          doi = {10.1093/mnras/stz2362},
archivePrefix = {arXiv},
       eprint = {1905.11516},
 primaryClass = {astro-ph.HE},
       adsurl = {https://ui.adsabs.harvard.edu/abs/2019MNRAS.489.3591P}
}

@article{Villar_2021,
doi = {10.3847/1538-4365/ac0893},
url = {https://doi.org/10.3847/1538-4365/ac0893},
year = {2021},
month = {aug},
publisher = {The American Astronomical Society},
volume = {255},
number = {2},
pages = {24},
author = {Villar, V. Ashley and Cranmer, Miles and Berger, Edo and Contardo, Gabriella and Ho, Shirley and Hosseinzadeh, Griffin and Lin, Joshua Yao-Yu},
title = {A Deep-learning Approach for Live Anomaly Detection of Extragalactic Transients},
journal = {The Astrophysical Journal Supplement Series}
}

@article{Bellm_2019,
doi = {10.1088/1538-3873/aaecbe},
url = {https://doi.org/10.1088/1538-3873/aaecbe},
year = {2018},
month = {dec},
publisher = {The Astronomical Society of the Pacific},
volume = {131},
number = {995},
pages = {018002},
author = {Bellm, Eric C. and Kulkarni, Shrinivas R. and Graham, Matthew J. and Dekany, Richard and Smith, Roger M. and Riddle, Reed and Masci, Frank J. and Helou, George and Prince, Thomas A. and Adams, Scott M. and Barbarino, C. and Barlow, Tom and Bauer, James and Beck, Ron and Belicki, Justin and Biswas, Rahul and Blagorodnova, Nadejda and Bodewits, Dennis and Bolin, Bryce and Brinnel, Valery and Brooke, Tim and Bue, Brian and Bulla, Mattia and Burruss, Rick and Cenko, S. Bradley and Chang, Chan-Kao and Connolly, Andrew and Coughlin, Michael and Cromer, John and Cunningham, Virginia and De, Kishalay and Delacroix, Alex and Desai, Vandana and Duev, Dmitry A. and Eadie, Gwendolyn and Farnham, Tony L. and Feeney, Michael and Feindt, Ulrich and Flynn, David and Franckowiak, Anna and Frederick, S. and Fremling, C. and Gal-Yam, Avishay and Gezari, Suvi and Giomi, Matteo and Goldstein, Daniel A. and Golkhou, V. Zach and Goobar, Ariel and Groom, Steven and Hacopians, Eugean and Hale, David and Henning, John and Ho, Anna Y. Q. and Hover, David and Howell, Justin and Hung, Tiara and Huppenkothen, Daniela and Imel, David and Ip, Wing-Huen and Ivezić, Zeljko and Jackson, Edward and Jones, Lynne and Juric, Mario and Kasliwal, Mansi M. and Kaspi, S. and Kaye, Stephen and Kelley, Michael S. P. and Kowalski, Marek and Kramer, Emily and Kupfer, Thomas and Landry, Walter and Laher, Russ R. and Lee, Chien-De and Lin, Hsing Wen and Lin, Zhong-Yi and Lunnan, Ragnhild and Giomi, Matteo and Mahabal, Ashish and Mao, Peter and Miller, Adam A. and Monkewitz, Serge and Murphy, Patrick and Ngeow, Chow-Choong and Nordin, Jakob and Nugent, Peter and Ofek, Eran and Patterson, Maria T. and Penprase, Bryan and Porter, Michael and Rauch, Ludwig and Rebbapragada, Umaa and Reiley, Dan and Rigault, Mickael and Rodriguez, Hector and Roestel, Jan van and Rusholme, Ben and Santen, Jakob van and Schulze, S. and Shupe, David L. and Singer, Leo P. and Soumagnac, Maayane T. and Stein, Robert and Surace, Jason and Sollerman, Jesper and Szkody, Paula and Taddia, F. and Terek, Scott and Van Sistine, Angela and van Velzen, Sjoert and Vestrand, W. Thomas and Walters, Richard and Ward, Charlotte and Ye, Quan-Zhi and Yu, Po-Chieh and Yan, Lin and Zolkower, Jeffry},
title = {The Zwicky Transient Facility: System Overview, Performance, and First Results},
journal = {Publications of the Astronomical Society of the Pacific}
}

\end{document}